\documentclass[structabstract]{aa}  
\usepackage[dvips]{graphicx}
\usepackage{txfonts}
\usepackage{natbib}
\usepackage{version}

\begin{document}
\setlength{\topmargin}{0.1cm}
     \title{The characterisation of irregularly-shaped particles:}

   \subtitle{a re-consideration of finite-sized, `porous' and `fractal' grains\\[0.2cm] 
{\small Published: A\&A, 528, A98 (2011) -- DOI: 10.1051/0004-6361/201015968}}

   \author{A. P. Jones
          \inst{}
          }

   \institute{IInstitut d'Astrophysique Spatiale, CNRS/Universit\'e Paris Sud, UMR\,8617, Universit\'e Paris-Saclay, Orsay F-91405, France\\
               \email{ Anthony.Jones@ias.u-psud.fr}
             }

   \date{Received October 2010 ; accepted 15/01/2011}

 
  \abstract
   {A porous and/or fractal description can generally be applied where particles have undergone coagulation into aggregates. 
   }
   {To characterise finite-sized, `porous' and `fractal' particles and to understand the possible limitations of these descriptions.
   }
   {We use simple structure, lattice and network considerations to determine the structural properties of irregular particles. }
   {We find that, for finite-sized aggregates, the terms porosity and fractal dimension may be of limited usefulness and  show with some critical and limiting assumptions, that  `highly-porous' aggregates (porosity $\gtrsim 80$\%) may not be `constructable'. 
    We also investigate their effective cross-sections using a simple `cubic' model.   
      }
   {In place of the terms porosity and fractal dimension, for finite-sized aggregates, we propose the readily-determinable quantities of inflation,   $I$ (a measure of the solid filling factor and size), and dimensionality, $D$ (a measure of the shape).
   These terms can be applied to characterise any form of particle, be it an irregular, homogeneous solid or a highly-extended aggregate.}


   \maketitle
%

\section{Introduction}

Interstellar, interplanetary or cometary dust particles are generally modelled as spheres, or collections of mono- or multi-disperse spheres. Such approaches are clearly rather idealised starting points from which to model and to perform analogue experiments aimed at understanding the obviously much more complex and diverse forms of dust in astrophysical media. 
For example, many of the collected interplanetary dust particles, or IDPs, show complex, extended and open structures comprised of aggregates of irregular sub-grains with radii typically of the order of hundreds of nanometres \citep[e.g.,][]{brownlee78,rietmeijer98}. Such types of structures are to be expected in coagulated grains in interstellar and proto-planetary disc particles \citep[e.g.,][]{ossenkopf93,ormel07,ormel09} and are seen in soot particle experiments and simulations \cite[e.g.,][and references therein]{dobbins91,koylu95,mackowski06}. 

Highly-porous grains, of apparently arbitrarily-large porosity, can be constructed mathematically but can they exist physically, i.e., can we build a physical 3D model of them and do such particles bear any resemblance to `real' interstellar grains? Such particles also pose the question as to how one should define porosity in these cases.

In this paper we discuss the issues of aggregate/irregular grain porosity and fractal structure and attempt to define characteristic properties for aggregate grains, composed of single-sized spherical constituent grains (also known as sub-grains or primary particles) that can be determined using simple measures of the aggregate geometry. We show that this approach can be extended to any irregularly-shaped particle.


\section{Porosity matters}

The porosity, $P$, of a particle is usually defined as, 
\begin{equation}
P = \left( \frac{V_v}{V_t} \right) = \left( 1 - \frac{V_s}{V_t} \right), 
\end{equation}
where $V_v$ and $V_s$ are the volumes of vacuum and of the solid matter making up the particle, and $V_t$ is the total volume of the particle within some defined surface. Note that this definition of porosity is independent of the form of the particles, which can be of any arbitrary shape, and is only defined for $0 \leq P < 1$. 
The problem with this definition of porosity is that it is often difficult to define a physically-reasonable surface that delimits the particle in an unequivocal way. 
Clearly, a measure of a particle's porosity requires an {\it unambiguous} definition of the reference volume, which is in many, if not most, cases not possible. Thus, the porosity of very extended particles can be hard to define and it may not make much sense to talk of `highly porous'  particles. It therefore seems that a better measure or a more open-ended definition of the nature of coagulated grains is needed. 
\cite{ormel07} and \cite{ormel09} have considered the `porosity' of coagulated grains in terms of an enlargement parameter and a geometrical filling factor approach to determine the particle cross-sections, respectively. We return to a discussion of these ideas later.


\section{Fractal matters}

We recall that a fractal structure has the property of self-similarity and that the constituent parts of its structure can be considered a reduced copy of the whole, i.e., the details of the structure are essentially scale- or reference point-invariant.
The characteristic of a fractal structure is its R\'enyi, Hausdorff or packing dimension (i.e., the fractal dimension). 
In a rather general way the fractal dimension, $D_f$, of a system can be defined as follows:
\begin{equation}
N(r) \ \ {\rm or} \ \ M(r) \propto r^{D_f}, \ \ \ \ \ \ 1 \leq D_f \leq 3
\end{equation}
where $N(r)$ and $M(r)$ are the number and mass, respectively, of particles within a defined radius, $r$, from a given reference point. This definition of the fractal dimension, or what might perhaps be called the `fractality' of the system, is determined by the spatial distribution of the structural units existing within a defined spherical volume of a much larger entity. For an infinitesimally-reducible structure, such as the Mandelbrot set, this clearly poses no problem. However, for a finite sized, porous particle, which does not exist beyond this assumed spherical volume and does not reproduce itself at ever smaller size-scales, the idea of a scale-invariant fractal dimension no longer holds.
Interstellar grains therefore cannot be fractal in the true sense of the definition of fractal entities because their structures are not scale-invariant, i.e., a shift in the reference point will not always yield the same fractal dimension. For example, a finite piece of `chicken wire' (essentially a 2D hexagonal grid or mesh) might be considered to have a fractal dimension of two globally but one locally when one descends to the level of the wire and no longer sees the larger mesh.  Thus, the use of the term fractal, when referring to finite-sized particles, can be rather misleading. It would seem that a better description for irregular interstellar grains ought to be possible and should be rather general. 

As we will show low $D_f$ particles cannot be `porous' because this leads to a loss of continuity in the structure. Thus, the `fractal' nature of a structure is coupled to its `porosity' and the two properties therefore cannot be independently varied, as already discussed in some detail by \cite{mandelbrot82}.


\section{Sparse lattices}

By sparse lattice we here mean stacked constituents on a regular lattice (e.g., cubic, hexagonal or face-centred cubic) in which not every lattice site is necessarily occupied.


\subsection{Simple cubic lattices}
\label{sect_simple_cubic}

A simple cubic lattice, i.e., a lattice composed of equi-dimensional regular cubic constituents, can be packed together with 100\% efficiency leaving no intervening, unfilled space. This lattice is clearly not of relevance to the structure of interstellar grains but it does, nevertheless, provide some useful intuitive insight into the nature of porous grain structures.

In a simple cubic lattice each site shares 6 faces, 12 edges and 8 corners with neighbouring lattice sites. If we now randomly remove lattice sites we can look at how the number of neighbouring sites depends on the porosity. For illustrative purposes we consider constituent cubes of edge length $l$ stacked into a cubic particle of edge length $nl$ (i.e., total number of lattice sites  $= n^3$). A cubic particle of porosity $P$ then contains $(1-P) n^3$ occupied sites and the average number of occupied adjacent sites (ignoring the macroscopic particle outer edges and surfaces), which we call the coordination number for a given porosity, $C_P$, is simply $(1-P)C$, where $C$ is the maximum site coordination number under consideration. For a cubic lattice, $C$ may take the values 6, 18 or 26, depending on the hierarchy of the coordination that we allow, i.e., sites sharing only faces ($C = 6$), sites sharing faces and edges ($C = 6+12 = 18$) sites sharing faces, edges and corners ($C = 6+12+8 = 26$). We then have that
\begin{equation}
C_P = (1-P) C, {\rm where}\ C = 6,\ 18\ {\rm or}\ 26.
\end{equation}

We are now interested in the limit where the structure is just continuous and, therefore, can be physically constructed as a single entity. Two limits are of interest here: 
\begin{enumerate}
\item The {\it dimer limit} where each site has only one occupied neighbouring site and the system therefore consists only of dimers. Such a `structure' is clearly not macroscopic, does not `hold up' and is not `constructable'.
\item The {\it continuity limit}, which is somewhat akin to the {\it percolation threshold} for a lattice, where the structure is just bound into a single, `one-dimensional' unit by inter-site contacts or bonds. 
\end{enumerate}
The first case is trivial and $C_P =1$. The second case is of interest in determining whether a given structure is `constructable' and in this case $C_P$ is just equal to 2, which ensures that the network can, minimally, completely inter-link. In this case each site has two filled neighbouring sites and the network can, in theory at least, propagate over the whole considered volume as a 1D `chain'. Although, in the latter case the structure would resemble something like a ball of wire, i.e., a one-dimensional construct warped into a three dimensional space, which is not a fractal structure.
 
For the simple cubic lattice, or in fact for any lattice, the relevant critical porosities $P_{\rm crit}$ for a given case are given by
\begin{equation}
P_{\rm crit} = \left( 1 - \frac{C_P}{C} \right).
\end{equation}
For the {\it dimer limit}, i.e., $C_P = 1$, we then have that $P_{\rm crit} = \frac{5}{6},\ \frac{17}{18}\ {\rm and}\ \frac{25}{26}$, respectively, for the $C =$ 6, 18 and 26 maximum coordination cases.
For the more interesting {\it continuity limit} case, where $C_P = 2$,  we have $P_{\rm crit} = \frac{4}{6},\ \frac{16}{18}\ {\rm and}\ \frac{24}{26}$, respectively, for the $C =$ 6, 18 and 26 maximum coordination cases. In the latter case, i.e. $C = 26$, allowing a structure to coordinate only through the corner points is perhaps a rather extreme situation. The equivalent porosities for the three cases are $\approx$ 66, 89 and 92\%. For comparison, percolation theory, see for example \cite{percolation}, predicts site and bond percolation thresholds of 0.3116 and 0.2488, respectively, which would be equivalent to `critical porosities' of $\approx$ 69 and 75\%. Our simple model therefore indicates values roughly consistent with percolation theory.


\subsection{Close-packed lattices}

We now turn our attention to the perhaps more astrophysically-relevant, but still idealised, case of close-packed lattices of spherical entities. In this case we note that the lattice positions are fixed and that the constituent `monomers' cannot take arbitrary positions. Nevertheless, this is a useful starting point.

By simple close-packed lattices we refer to lattices composed of regular, equal-sized spheres packed as efficiently as possible, i.e., face-centred cubic (fcc) and hexagonal close packing (hcp), which are identical for our purposes. However, for the simplicity of the presentation, we will only consider the fcc case for simple solid spheres. For identical size and composition spheres, the fcc lattice unit cell can be considered as a cube where each face of the cube shares a sphere with one adjacent unit cell and each corner of the cube shares a sphere with seven other neighbouring unit cells. It then follows that each fcc unit cell contains $(6 \times \frac{1}{2}) + ( 8 \times \frac{1}{8}) = 4$ spheres within a cube of side length $\sqrt{2} d$ where $d$ is the diameter of the sphere. The volume of the cube is then
\begin{equation}
V_c = 2 \sqrt{2} d^3
\end{equation}
and the volume occupied by the solid matter of the spheres is
\begin{equation}
V_s = 4 \times \frac{4}{3} \pi \left(\frac{d}{2}\right)^3 = \frac{2}{3} \pi d^3. 
\end{equation}
The porosity of such a vacancy- or defect-riddled fcc structure is then given by
\begin{equation}
P_{\rm fcc} = \left( 1 - \frac{V_s}{V_c} \right) =  \left( 1 - \frac{\frac{2}{3} \pi d^3}{2 \sqrt{2} \ d^3} \right) = \left( 1 - \frac{\pi}{3 \sqrt{2}} \right),
\end{equation}
i.e., a porosity of 26\%, which is equivalent a solid filling factor of $F_{\rm fcc} = \frac{ \pi }{ 3 \sqrt{2} } = 0.74$, which appears to be the most efficient lattice packing possible for equal-sized spheres. The same value holds for the hcp lattice. For the simple cubic packing of spheres it can similarly be shown that the filling factor is $ F_{\rm cubic} = \frac{ \pi }{ 6 } = 0.52$. These results can be found in standard textbooks on crystallography and the solid state.

We now consider a spherical volume of radius $A$ and volume $V_g$, perhaps equivalent to an aggregate interstellar grain, composed of spherical sub-grains, of volume $V_{\rm sg}$, packed on a fcc lattice.  
The maximum number of sub-grains, $N_{\rm max}$, within the total grain volume is then
\begin{equation}
N_{\rm max} = \frac {  V_{\rm g} F_{\rm fcc}}{ V_{\rm sg}} 
= \frac{ \frac{4}{3} \pi A^3 F_{\rm fcc} }{ \frac{4}{3} \pi \left(\frac{d}{2}\right)^3 } 
= 8 F_{\rm fcc} \left( \frac{A}{d} \right)^3
= \frac{ 8 \pi }{ 3 \sqrt{2} } \left( \frac{A}{d} \right)^3.
\end{equation}
We now use the above results to study the effects of introducing porosity into the structure by removing some of the sub-grains from the lattice. If $N$ is the total number of sub-grains within the particle or aggregate ($N \leq N_{\rm max}$) then, following from the definition above (with $V_{\rm sg} = \frac{4}{3} \pi (\frac{d}{2})^3 = \frac{\pi}{6}d^3$), the porosity of the particle is given by
\begin{equation}
P =  \left( 1 - \frac{ N \frac{\pi}{6} d^3}{ \frac{4}{3} \pi A^3  \ F_{\rm fcc} } \right)
= \left[ 1 - N \left( \frac{d}{A} \right)^3 \frac{3 \sqrt{2}}{8 \pi} \right]
= \left( 1 - \frac{N}{N_{\rm max}} \right), 
\end{equation}
or, equivalently, the number of sub-grains for a given porosity, $N_P$, is given by
\begin{equation}
N_P = (1-P) \ 8 F_{\rm fcc} \left( \frac{A}{d} \right)^3 
= (1-P) \ \frac{8 \pi}{3 \sqrt{2}} \left( \frac{A}{d} \right)^3.
\end{equation}
where we recover the value $N_{\rm max}$ when $P = 0$. 

The average sub-grain coordination number for our porous lattice, which can be expressed in the same way as for the simple cubic lattice discussed above, is simply the maximum coordination number, $C$ (12 for fcc and hcp), multiplied by the fraction of filled sites $(1-P)$, i.e., $C_P = (1-P) C$. In this case the {\it dimer limit} and {\it continuity limit} critical porosities, $P_{\rm crit}$, are  $\frac{11}{12}$ and $\frac{5}{6}$, i.e., 92\% and 83\%, respectively. Thus, it appears that, for this idealised close-packed lattice (fcc or hcp) model, there is an upper limit to the porosity of a particle and that this limit is $\approx 80\%$. For the cubic lattice the equivalent limiting porosity is $\approx 70\%$. Note that these limits are given by $(1-[2/C])$ and are therefore independent of the monomer size. 

However, remember that this limiting porosity strongly depends upon the above, and widely used, definition or porosity and the fixed lattice positions  for our particle subgrains. It also depends on the fact that we have assumed that the sub-grains are `evenly' distributed throughout the volume $\frac{4}{3} \pi A^3$. We now look at the nature and definition of porosity more closely.


\section{Sparse networks}

By sparse network we here mean contiguous, irregularly-connected, spherical constituents that do not lie on a regular lattice.


\subsection{General packing of spherical grains}

We consider a mono-disperse collection of connected spherical sub-grains in an arbitrarily-shaped, `porous' particle that can be defined by its three semi-major axes ($a, b$ and $c$), as per an ellipsoid, with an `enclosing' volume, $V_{\rm e} = \frac{4}{3} \pi a b c$. The maximum number of sub-grains of diameter $d$ within this assumed volume is given by
\begin{equation}
N_{\rm max} = \frac {  V_{\rm e} \phi }{ V_{\rm sg}} 
= \frac{ \frac{4}{3} \pi \ a b c \ \phi }{ \frac{4}{3} \pi \left(\frac{d}{2}\right)^3 } 
= 8 \phi  \left( \frac{a b c}{d^3} \right), 
\end{equation}
similar to that defined for a spherical particle above but where we have now replaced the maximum fcc lattice packing efficiency for mono-disperse, spherical particles, $F_{\rm fcc} = \pi / (3 \sqrt{2})$, by a more-generalised maximum packing efficiency $\phi$. The porosity of the particle is given by
\begin{equation}
P =  \left( 1 - \frac{ N \ \frac{\pi}{6} d^3}{ \frac{4}{3} \pi \ a b c  \ \phi } \right)
= \left[ 1 - \frac{N}{8 \phi} \left( \frac{d^3}{a b c} \right) \right]
= \left( 1 - \frac{N}{N_{\rm max}} \right), 
\end{equation}
and the number of sub-grains for a given porosity, $N_P$, is 
\begin{equation}
N_P = (1-P) \ 8 \phi \left( \frac{a b c}{d^3} \right) .
\end{equation}
We can now consider a more generalised definition of the semi-major axes, $a, b$ and $c$, of our porous particle that is independent of the filling factor. We normalise the three principal axes of the particle (i.e., $2a$, $2b$ and $2c$) by the monomer size, $d$. If we assume that these axes are parallel to the cartesian axes $x, y$ and $z$, and that $N_x = 2 a/d,\ \ N_y = 2 b/d$ and $N_z = 2 c/d$, then the volume within the encompassing surface is
\begin{equation}
V_e = \frac{4}{3} \pi \times \frac{N_x \ d}{2} \times \frac{N_y \ d}{2} \times \frac{N_z \ d}{2}
\ = \frac{\pi}{6} \  d^3 \ ( N_x \, N_y \, N_z).
\label{Venc}
\end{equation}
We can simplify the above expressions for $N_{\rm max}$, $P$ and $N_P$ to:
\begin{equation}
N_{\rm max} = \phi  \left(  N_x \, N_y \, N_z \right), 
\end{equation}
\begin{equation}
P = \left[ 1 - \frac{1}{\phi } \left( \frac{N}{N_x \, N_y \, N_z} \right) \right]
= \left( 1 - \frac{N}{N_{\rm max}} \right), 
\end{equation}
\begin{equation}
N_P = (1-P) \ \phi \ \left( N_x \, N_y \, N_z \right)
= (1-P) N_{\rm max}.
\end{equation}

\begin{figure}
\centering 
\includegraphics[width=8.5cm]{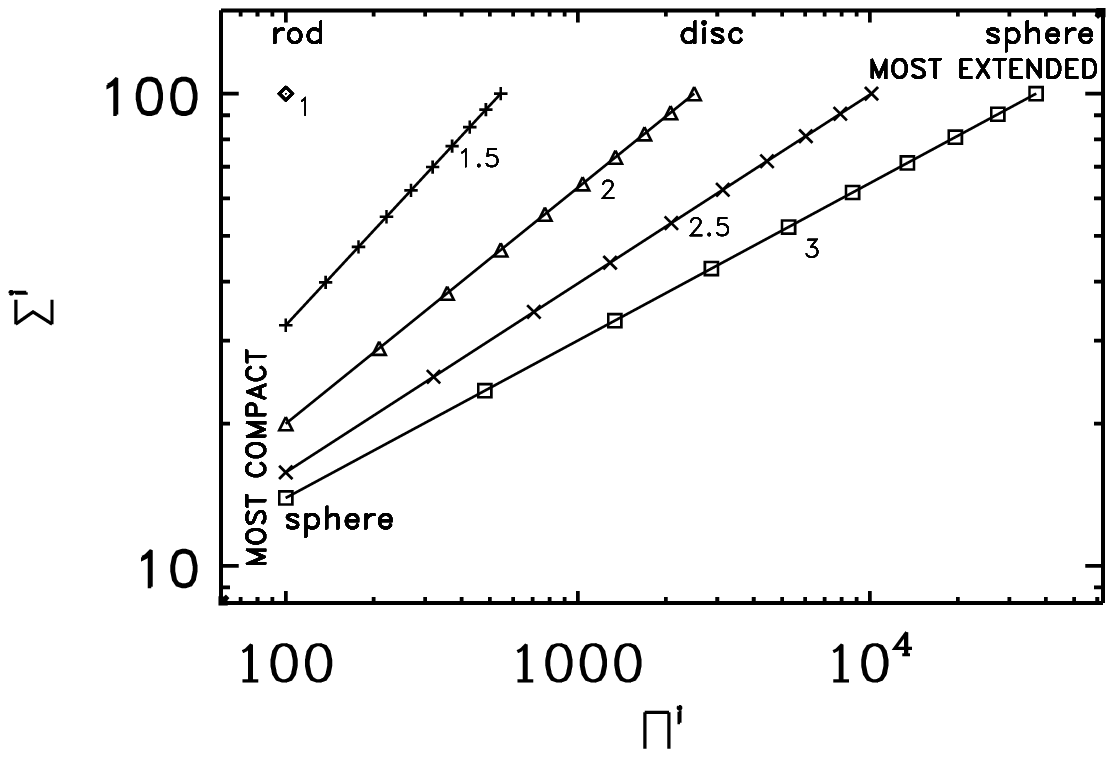}
\caption{Axis sum, $\Sigma^i$, vs. axis product, $\Pi^i$, plot for a particle containing 100 identical sub-particles as a function of the dimensionality (as indicated): $D=1$ (diamond), 1.5 (plus signs), 2 (triangles), 2.5 (crosses) and 3 (squares).  Compact particles are to the lower left and extended particles to the upper right. Note that no porosity is allowed for $D = 1$, rod-like particles.}
\label{fig_porous_fractal}
\end{figure}

The key parameters in defining our arbitrary particle are now the number of constituent monomers, $N$, and its normalised dimensions, $N_x, N_y$ and $ N_z$ or more generally $N_i$ where $i = x, y$ or $z$. The nature of the particle can be characterised, independent of porosity or fractal measures, through the sum, $\Sigma^i$,  and the product, $\Pi^i$, of its dimensions, i.e.,
\begin{equation}
\Sigma^i = \sum^i N_i = ( N_x + N_y + N_z )
\end{equation}
\begin{equation}
\Pi ^i = \prod^i N_i =  ( N_x \times N_y \times N_z ).
\end{equation}
Table~\ref{tab1} shows the values of $\Sigma^i$  and $\Pi^i$ for the most compact and the most extended particles possible. For highly porous `disc-like' and `sphere-like' distributions this corresponds to linear structures located along 2 and 3 orthogonal axes, respectively. Note that a one-dimensional or rod-like assemblage of spheres can never be porous since any `vacancy' cuts the structure into smaller but separated particles. Thus, the maximum possible particle dimension is $N d$ and the minimum particle dimension is that of the most compact spherical particle possible, i.e., $\phi N^{1/3} d$. Fig.~\ref{fig_porous_fractal} shows the relationship between $\Sigma^i$  and $\Pi^i$ as a function of the particle dimensionality, which is defined below.

\begin{table}
\caption{$\Sigma^i$  and $\Pi^i$ for compact and highly extended particles.}             
\label{tab1}      
\centering                          
\begin{tabular}{llcc }        
\hline\hline \\[-0.25 cm]                 
Shape & dimensions & $\Sigma^i$ & $\Pi^i$ \\    
\hline \\[-0.25 cm]                        

\multicolumn{4}{l}{Most compact}  \\
sphere & $N_x = N_y = N_z = N^{1/3}       $ & $ 3 N^{1/3} $ & $ N $ \\      
disc     & $N_x = 1 \ll N_y = N_z = N^{1/2} $ & $ 2 N^{1/2} $ & $ N $ \\
rod      & $N_x = N_y =1 \ll N_z = N           $ & $    N           $ & $ N $ \\
\hline \\[-0.25 cm]
\multicolumn{4}{l}{Most extended}  \\
sphere & $N_x = N_y = N_z = N/3 $ & $N$ & $\frac{1}{9}N^3$ \\
disc      & $N_x = 1 \ll N_y = N_z = N/2 $ & $N$ & $\frac{1}{4}N^2$ \\
rod       & $N_x = N_y = 1 \ll N_z = N   $ & $N$ & $N$ \\
\hline \\[-0.25 cm]                                   
\end{tabular}
\end{table}

We can see from Table~\ref{tab1} that for the most compact particles $\Sigma^i \leq N$ and $\Pi^i = N$ and for the most extended particles $\Sigma^i = N$ and $\Pi^i \geq N$. If we wish we can consider the porosity, $P$, of these particles and re-write $N_{\rm max}$, $P$ and $N_P$ as:
\begin{equation}
N_{\rm max} = \phi \ \Pi^i, 
\end{equation}
\begin{equation}
P = \left( 1 - \frac{N}{\phi \ \Pi^i} \right), 
\end{equation}
\begin{equation}
N_P = (1-P) \ \phi \ \Pi^i.
\end{equation}
However, the advantage of using the $\Sigma^i$  and $\Pi^i$ characterisation is that we have a system that can generically describe an arbitrarily-extended particle without the need to define its porosity or its `fractality'.


\subsection{Dimensionality and inflation}

With a view to generalising the notion of grain `porosity', or better characterising extended particles, we consider a more open-ended definition than simply the particle porosity, as did \cite{ormel07} and \cite{ormel09}. \cite{ormel07} considered grains in terms of an enlargement parameter, the ratio of the extended volume to the most compact volume. Whereas  \cite{ormel09} used a geometrical filling factor approach to determine the particle cross-sections. 
We now consider the idea of particle `inflation' beyond the most compact forms possible. In this case we define the inflation, $I$, for a given particle as:
\begin{equation}
I = \frac{\frac{1}{6} \pi ( N_x \times N_y \times N_z )d^3}{\frac{1}{6} \pi N d^3 \phi^{-1}}
= \frac{\phi ( N_x \times N_y \times N_z )}{N} = \frac{\phi \Pi^i}{N} = \frac{N_{\rm max}}{N}.
\end{equation}
Inflation is thus, equivalently, the ratio of the volume of the enclosing ellipsoid ($\frac{1}{6} \pi \, N_x N_y N_z \, d^3$, Eq.~\ref{Venc}) to the (minimum) volume of the solid matter in the aggregate ($\frac{4}{3} \pi \, [N^{\frac{1}{3}}]^3 \, (d/2)^3 \ \phi^{-1} = \frac{1}{6} \pi \, N \, d^3 \ \phi^{-1}$), where $\phi$ is the maximum packing efficiency of the sub-grains in the minimum volume. 
For cubic particles (on a cubic grid, see \S~\ref{sect_simple_cubic}), where $\phi =1$, we have $(\frac{1}{6} \pi \, N_x N_y N_z \, d^3)/(N d^3) = (\frac{1}{6} \pi \, N_x N_y N_z )/N$. Note that with this definition an inflation factor $I = 10$ could be considered akin to a porosity of 90\%. 

The inflation, $I$, for any particle has a minimum value of unity and a maximum value of $\approx N^2/27$ (for unit packing efficiency, i.e., $\phi =1$), which is determined by the most-extended  `spherical' particle possible. This hypothetical, and rather physically unrealistic, structure has an equal number of sub-particles ($N/3$) arranged along the three orthogonal, cartesian axes. This corresponds to the extreme upper right part of Fig.~\ref{fig_porous_fractal}. For a particle with a dimensionality of 2 the maximum value of $I$ is $\approx N/4$. We can see that with this definition the increase in the particle size by a factor of $f$ translates to an inflation factor of $f^3$, provided that the condition $( N_x + N_y + N_z ) \leq N$ is fulfilled.
Note that the largest values of $I$ are to be found for the most `spherical' extended particles where the product, $\Pi^i$, is a maximum for a given sum, $\Sigma^i$.

We now define the dimensionality, $D$, of any given extended particle in terms of its maximum-dimension-normalised sum $\Sigma^i$:
\begin{equation}
D = \frac{\Sigma^i}{{\rm max}\{ N_x, N_y, N_z \}} = \frac{( N_x + N_y + N_z )}{N_{L+}}, 
\end{equation}
where $N_{L+} = {\rm max}\{ N_x, N_y, N_z \}$ and indicates that we take the longest dimension of the particle, as the normalisation length. 

It is clearly possible to extend this definition of $D$ to the general case of any arbitrary-shaped particle, independent of whether it is an aggregate or a highly irregular single `lump', i.e.,
\begin{equation}
D = \frac{L_x + L_y + L_z}{{\rm max}\{ L_x, L_y, L_z \}}, 
\end{equation}
where $L_i$ is the particle size along its $i^{\rm th}$ axis ($x$, $y$ or $z$). As an illustrative example, in Table~\ref{shapes} we show the dimensionality and inflation factors for several regular solids, where the value $L_x$ ($L_z$) is, generally, the maximum (minimum) dimension of the regular solid. Note that in the case of the cube the longest dimension, $L_x$, is the cube diagonal, and not an edge, and that the other two dimensions are then taken orthogonal to this direction.

\begin{table*}
\caption{The dimensionality, $D$, and inflation, $I$, for regular geometrical solids. $a$ and $l$ are the relevant radius and side/edge lengths, respectively.}             
\label{shapes}      
\centering                          
\begin{tabular}{lcccccc }        
\hline\hline \\[-0.25 cm]                 
Solid & $L_x$  & $L_y$ & $L_z$ & volume & $D$ & $I$  \\    
\hline \\[-0.05 cm]                        

sphere         & 2$a$                      & 2$a$                                    & 2$a$                   & $\frac{4}{3} \pi a^3$       & 3      & 1                       \\[0.2cm]
octahedron  & 2$a$                      & 2$a$                                    & 2$a$                   & $\frac{1}{3} \sqrt{2} l^3$ & 3      & $\pi$                 \\[0.2cm]
cube            & $\sqrt{3} l$             & $\frac{2\sqrt{2}}{\sqrt{3}} l$ & $\sqrt{2} l$           & $l^3$                              & 2.76 & $\frac{2}{3} \pi$ \\[0.2cm]
tetrahedron & $\frac{1}{\sqrt{2}}l$ & $l$                                       & $l$                       & $\frac{\sqrt{2}}{12} l^3$          & 2.71 & $\pi$                  \\[0.2cm]
cylinder       &  $\sqrt{4a^2+l^2}$   & $4al/ \left[ \sqrt{4a^2+l^2} \right]$  & 2$a$ & $\pi a^2l$ & $1+ \left( 4al/ \left[ 2a^2+l^2 \right] \right)+ \left( 2a/  \left[ \sqrt{4a^2+l^2} \right] \right)$ & $\frac{4}{3}$ \\[0.2cm]
\ \ \ disc ($a \gg l$)  & $\approx$2$a$ & $\rightarrow 0$ & 2$a$  & $\pi a^2l$  & $\approx$2 & $\frac{4}{3}$ \\[0.2cm]
\ \ \ rod ($a \ll l$)     & $\approx$$l$    & $\rightarrow 0$ & 2$a$  & $\pi a^2l$  & $\approx$1 & $\frac{4}{3}$ \\[0.2cm]
\hline \\[-0.25 cm]
\end{tabular}
\end{table*}

For our general case we can define a minimum and a maximum particle dimension, $N_{L-}$ and $N_{L+}$, respectively, defined by  the minimum volume (spherical) close-packed particle, of radius $a$, and the maximum linear chain length. These extreme particle dimensions are given, for the fcc close-packed and the more general case, by
\begin{equation}
N_{L-} = 2a = F_{\rm fcc} \ N^\frac{1}{3} d = \frac{\pi}{3 \sqrt{2}} \ N^\frac{1}{3} d = \phi \ N^\frac{1}{3} d 
\end{equation}
and
\begin{equation}
N_{L+} = N \ d.
\end{equation}
Thus, all possible particles must have sizes, $L$ (expressed in terms of the sub-unit dimension $d$), within the range 
\begin{equation}
\phi \ N^\frac{1}{3}  < \frac{L}{d} < N ,
\end{equation}
where the lower limit can be equated with a fractal dimension $D_f \approx 3$ and the upper limit with $D_f \approx 1$. However, the particles in question are not necessarily `truly' fractal in nature.
In fact, this definition of the quantity dimensionality ($1 \leq D \leq 3$) can indeed be applied to any particle, whether solid and homogeneous ($I =1$), particles with concavities and protrusions or porous aggregates ($I > 1$). Additionally, the concept of inflation can be applied to any particle because it is just the volume of the enclosing ellipsoid, that can directly determined from the particle dimensions, divided by the minimum volume of the solid matter ($V_s/\phi$, where $\phi$ is set to the maximum packing efficiency and is set to 1 for a solid particle).

Clearly any given, finite-sized particle can only have dimensionalities in the range 1 to 3. The dimensionality, $D$, when coupled to $I$, can perhaps be considered as somewhat analogous to the fractal dimension of a structure and as a representation of the spatial arrangement. 

\begin{figure}
\centering 
\includegraphics[width=8.5cm]{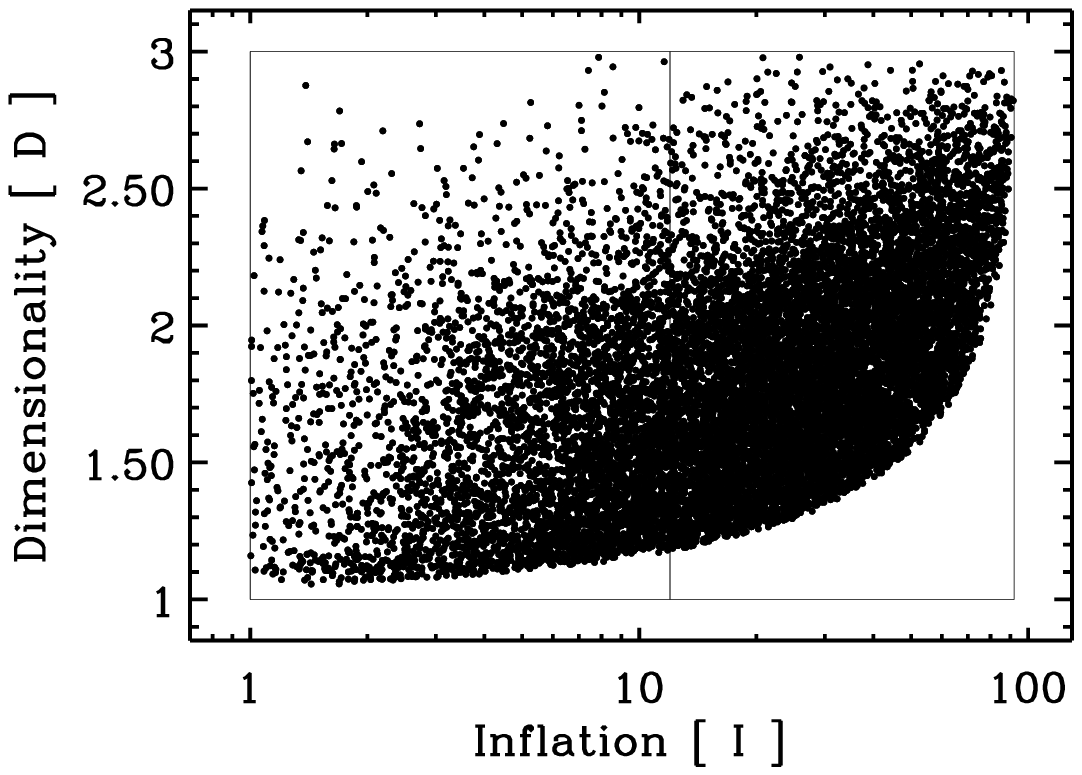}
\caption{Dimensionality, $D$, versus inflation, $I$, plot for particles containing 50 equal-sized constituents for $\approx 15,000$ randomly-generated structures 
{\bf 
(with $\phi = 1$). 
}
The inset box shows the limits on $D$ ($1 \leq D \leq 3$) and $I$ ($1 \leq I \leq N^2/27$), and the middle vertical line shows the limit for an idealised 2D particle, i.e., $I = N/4$.}
\label{fig_sum_prod_plot}
\end{figure}

The concepts of porosity and fractal structure (`fractality') for any given extended and `constructable' particle can now re-posed in terms of the related and estimable quantities of inflation ($I$) and dimensionality ($D$). In Fig.~\ref{fig_sum_prod_plot} we show a plot of $D \ vs. \ I$ for $\approx 15,000$ randomly-generated particles containing 50 identical sub-grains. 
For illustrative purposes we have assumed maximum packing efficiency (i.e., $\phi =1$). 
These aggregates were generated by imposing the conditions: $1 \leq N_i \leq N$ (where $i = x$, $y$ or $z$), $\Sigma^i \leq N$ and $\Pi^i \geq N$. From this figure it appears that particles created in this way are dominated by dimensionalities of the order of $\approx 1.2-2.3$ and inflation factors of $\approx 5-50$. The lower cut-off to the data points in this figure is simply due to the relationship between the sum of the particle dimensions, which determines the maximum dimension and therefore $D$, and the product of the particle dimensions, which determines the volume and hence $I$.

\begin{figure}
\centering 
\includegraphics[width=8.5cm]{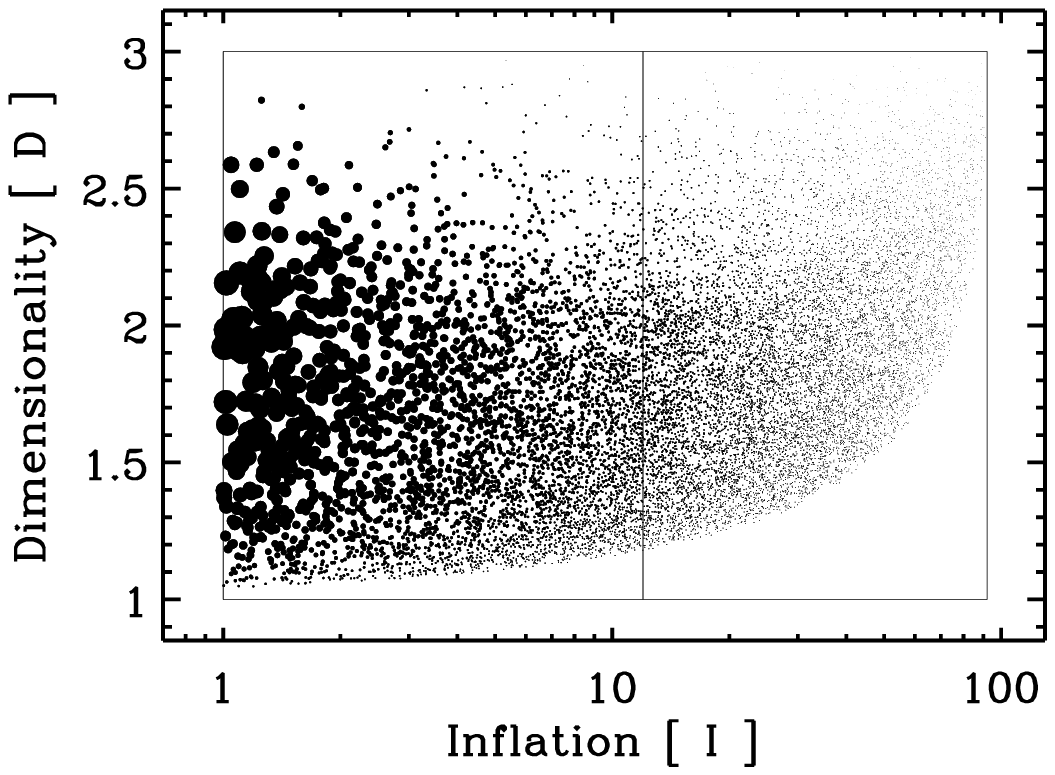}
\caption{Same as for Fig.~\ref{fig_sum_prod_plot} except that we have now volume- and shape-weighted the data point sizes by a factor $(1-[D-2]^2)/I$ (see text).}
\label{fig_sum_prod_plot_2}
\end{figure}

In Fig.~\ref{fig_sum_prod_plot} all particles are assumed to be equally `constructable'. However, given that the number of possible structural isomers for the given number of monomers (that we could perhaps call `structomers') must increase with the enclosing volume $V_e$, the more extended structures are given undue weight in Fig.~\ref{fig_sum_prod_plot}. This is indicated by the higher density of the data points for larger values of $I$ in this linear-log plot. Hence, and for illustrative purposes, in Fig.~\ref{fig_sum_prod_plot_2} we plot another simulation in which the size of the data points are volume-weighted by dividing by $I$ and shape-weighted by multiplying by $(1-[D_m-2]^2)$. For the shape-weighting we adopt an illustrative mean value of the dimensionality $D_m=2$, i.e., close to the typical fractal dimension (i.e., $D_f = 1.8$) seen in soot particle formation experiments. 
What Fig.~\ref{fig_sum_prod_plot_2} shows is that, assuming $D_m=2$, aggregate particles will probably have inflation factors $I < 10$. However, this will obviously require a more detailed investigation than presented here because the `structomer' distribution in the $D\ vs. I$ parameter space will depend upon the adopted particle construction protocol. 

Thus, what emerges from the above is that it is perhaps not the fractal dimension or even the porosity that counts for finite-sized, extended particles but the `dimensionality' of the structure coupled to its `inflation'. In Table~\ref{tab1} we can see that the `dimensionality' is represented in the exponents of $\Sigma^i$  and $\Pi^i$. 

In Appendix~\ref{appendix} we consider the implications of the above characterisation of irregular particles for the calculation of their cross-sections.

\section{Implications for astronomical dust studies}

Perhaps in no small part motivated by the then rather recent publication of the \cite{bohren83} book, early studies of `porous' and composite interstellar grains, which must implicitly be irregular in form, treated `porosity' as vacuum filling a given fraction of the particle `volume' but did not take into account effects arising from their irregular structure \citep[e.g.,][]{jones88,mathis89}. 
These early investigations of grain porosity, and those subsequent to them, adopted a variety of approaches to the problem, including: effective medium theories \citep[EMT, e.g.,][]{jones88,mathis89}, hollow spheres \citep[HS, e.g.,][]{jones88,min05}, multi-layered sphere models \citep[MLS, e.g.,][]{voshchinnikov05} and discrete dipole approximations such as DDA \citep[e.g.,][]{bazell90,perrin90,perrin91,henning93,wolff94}. 
These methods can be put into three general classes: 
1) implicit orientation-averaged (HS and MLS - spherical shell representations of an infinite number of randomly-orientated structures), 
2) optical property averaged methods (EMT - solid `dilution' by the addition of vacuum, e.g., see Appendix \ref{sect_EMT_averaged}), and 
3) fixed-shape methods where orientation- and structure-averaging is required (DDA). 
Comparison studies \citep[e.g.,][]{jones88,wolff94,wolff98,voshchinnikov05,shen08,shen09} reveal that the results of these very different methods actually compare rather well. However, the `averaged' approach EMT, HS and MLS methods, as applied to spheroidal porous grains, cannot account for the details of the scattering properties of porous/irregular particles. In this sense DDA-generated particle studies do a much better job in determining the scattering properties of porous particles \citep[e.g.,][]{shen08,shen09} because they take into account shape effects. 
Recently \cite[][and their earlier work cited therein]{skorov10} studied very large, fluffy aggregates ($N \lesssim 2000$) and conclude that their optical properties are sensitive to the aggregate size parameter, that characterisation by only $N$ is insufficient (i.e., more detailed descriptions are needed) and that polarisation data can provide a measure of the aggregate fluffiness. 

The concepts of inflation, $I$, and dimensionality, $D$, as defined here, can be applied to any arbitrary-shaped particle that is `constructable' with the DDA approach. They can also be applied to aggregates of spherical primary particles or monomers that can be analysed with the T-matrix method \citep[TMM, e.g.,][]{mackowski96} and the generalised multiparticle Mie theory method \citep[GMM, e.g.,][]{xu95}. All such particles are characterisable in terms of $I$ and $D$. 
The `porosity' of very open structures (i.e., `highly porous' particles) is not unambiguously determineable, for the reasons discussed in the earlier sections, and the `fractal dimension' description for aggregates consisting of a rather limited number of monomers or primary particles does not appear tenable \citep[see earlier sections and also][]{koehler11}. 

The $D$ and $I$ descriptors may be particularly useful for characterising highly elongated structures and more meaningful than is possible with a porous particle description. Elongated structures actually correspond to low values of $D$ and, perhaps rather surprisingly, also to low values of $I$. The low values of $I$ arise in this case because the enclosing ellipsoid surface lies close to most of the constituent sub-particles because of its `cigar' shape. 

In a porous aggregate the projected cross-section can be significantly less than that adopted in the `averaged' and `dilution' methods (e.g., EMT, HS and MLS) where the particle is assumed to be spherical and defined by a porosity-weighted, effective radius (e.g., see Appendix~\ref{sect_averaged}). In the determination of the optical properties these methods actually do a surprisingly good job.  
In Appendices \ref{sect_EMT_averaged} and \ref{sect_EMT_averaged_longwav} we present some aspects of the Bruggemann EMT methods and dust cross-sections at long wavelengths that could be useful in studies of inhomogeneous particle studies.


\subsection{Irregular particle mean projected cross-sections} 

We now use the methods developed earlier, to explore the relationship between $D$ and $I$, to estimate the projected cross-sections of randomly-constructed aggregates within an ellipsoidal volume. For this exploratory model we use 81 cubic primary particles (i.e., $N = 81$) on a cubic grid (see \S~\ref{sect_simple_cubic}), which have unit packing efficiency ($\phi = 1$) and a `spherical' compact form with cube centres that lie within a circumscribed sphere of diameter 5 times the length of the side of a cube. As before, the values of $D$ and $I$ for the particles are randomly generated. 
The centre of each face of the $N_x \times N_y \times N_x$ `box' that encompasses the ellipsoid is assigned a primary particle, and this used as a seed to link the faces with randomly-generated but sequentially-coordinated primary cubes. This ensures that the values of $D$ and $I$ so-generated are fulfilled for the particle in question. Extra cubes are added, randomly but also sequentially-coordinated to existing particle cubes, until the aggregate contains the required 81 primary particles. 
The coordination number, $m_i$, for each cubic element $i$ in the generated aggregate is calculated by summing the number of occupied nearest neighbour cubes. The mean coordination number for each generated aggregate is then $\bar{m} = \sum_i m_i / N$. The minimum condition for a constructable, single particle is the continuity limit for the cubic lattice, i.e., $\bar{m} \geqslant 2$ (see \S~\ref{sect_simple_cubic}) and aggregates with $\bar{m} < 2$ are therefore rejected. In constructing an aggregate we only allow neighbour-coordination by cube faces and edges (see \S~\ref{sect_simple_cubic}), which corresponds to the condition $2 \leqslant m_i \leqslant 18$. However, coordination numbers up to 26 are, in principle, possible because of the random filling method used. For the aggregates $\bar{m}$ is always greater than 2, and generally less than 4 for these sparse lattices. 

\begin{figure}
\centering 
\includegraphics[width=8.5cm]{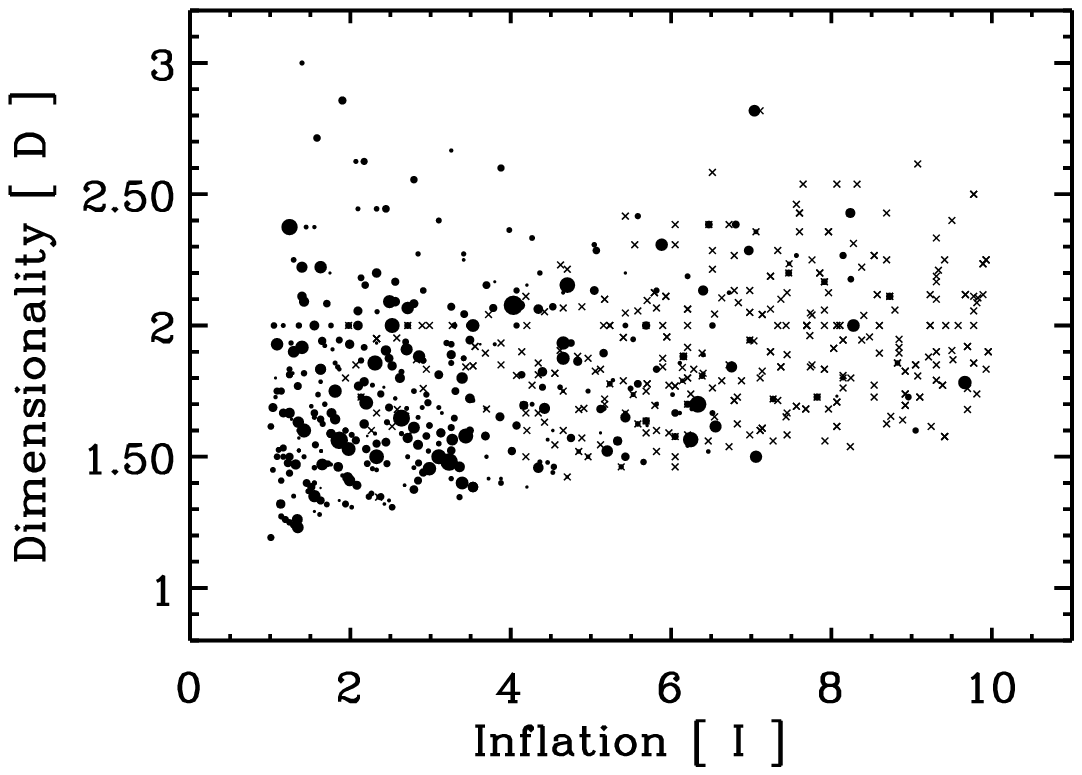}
\caption{The same as Fig.~\ref{fig_sum_prod_plot_2} but for $\approx 500$ aggregates (consisting of 81 cubic primary particles) and a more limited range of $I$ ($\le 10$). The symbols (filled circles) are size-scaled by $\bar{m}$. The crosses indicate `particles'  which are not constructable (see \S~\ref{sect_simple_cubic}). }
\label{fig_D_vs_I_limited}
\end{figure}

In Fig.~\ref{fig_D_vs_I_limited} we show a $D\ vs.\ I$ plot for a limited range of $I$ ($\le 10$). The symbols (filled circles) are size-scaled by the mean particle coordination number $\bar{m}$, which lies in the range $2 \le \bar{m} \lesssim 4$. The crosses indicate `particles' with $\bar{m} < 2$ that are not constructable (see \S~\ref{sect_simple_cubic}) and therefore eliminated from the cross-section analysis. Note that most of the rejected particles `envelope' the valid aggregates at higher values of $I$. 
The values of $\bar{m}$ are not systematically distributed in the $D\ vs.\ I$ space but appear to be somewhat random. 
Also note that the majority of the constructable aggregates have $I < 5$, which corresponds to porosities $\lesssim 80$\%. 
The curved series of data points for the valid aggregates, especially visible at the lower values of $I$ are due to the discreteness, in the limited range of compact aggregates for $N = 81$, and the coupling between $D$ and $I$ (see below).
%

\begin{figure}
\centering 
\includegraphics[width=8.5cm]{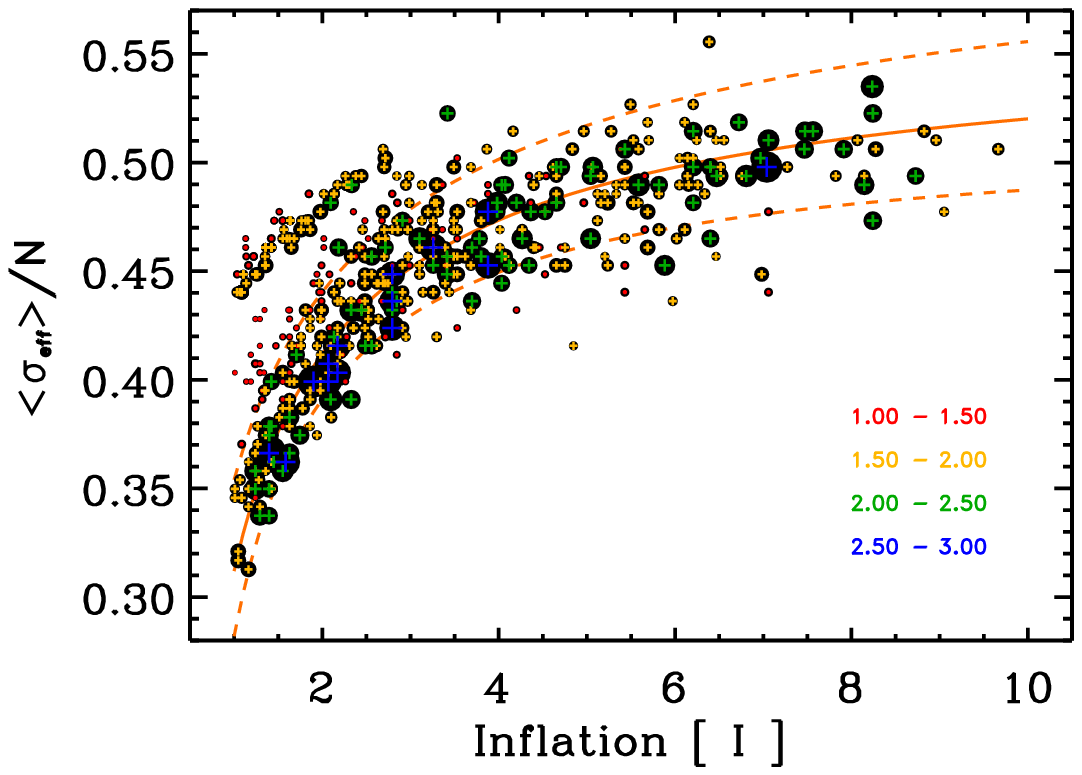}
\caption{The effective cross-section, normalised to the number of primary particles, $N$, for $\approx 500$ aggregates constructed from 81 cubic primary particles on a cubic grid.  
The symbols (black filled circles and coloured plus signs) are size-scaled by the particle dimensionality, $D$. The plus signs colour-code (see key at bottom right of the figure) the binned  dimensionality (red $D = 1-1.5$; yellow $D = 1.5-2$; green $D = 2-2.5$; blue $D = 2.5-2$). The orange lines shows some representative analytical fits (see text).}
\label{fig_sigma_xyz_vs_I}
\end{figure}

The projected cross-section for a given aggregate is calculated by viewing it along each of its axes, $x, y$ or $z$, and summing the number of occupied `pixels' in the plane projected on the other two axes, i.e., $\sigma_x(y,z)$, $\sigma_y(x,z)$ and $\sigma_z(x,y)$. The effective cross-section for the aggregate is then taken to be 
\begin{equation}
\langle \sigma_{\rm eff} \rangle = \frac{1}{3} \sum [\sigma_x(y,z) + \sigma_y(x,z) + \sigma_z(x,y)]. 
\label{eq_sigma_xyz }
\end{equation}
Fig.~\ref{fig_sigma_xyz_vs_I} shows the relationship between the aggregate projected cross-sections $\langle \sigma_{\rm eff} \rangle$, normalised by the number of primary particles $N$, and their inflation. The data points (filled circles) are scaled by the value of $D$ for the aggregate and are also colour-coded and binned by $D$  (plus signs): $1-1.5$ (red), $1.5-2$ (yellow), $2-2.5$ (green) and $2.5-3$ (blue). 
The solid orange line  in Fig.~\ref{fig_sigma_xyz_vs_I} shows an analytical fit, $\sigma_{\rm fit} = 0.52\,N\,(1.1+0.5 I^{-0.7})$, to the principle data points in this $ \langle \sigma_{\rm eff} \rangle\ vs. \ I$ plot. The upper [lower] dashed orange lines are for $\sigma_{\rm fit} = 0.59\,N\,(1.1+0.5 I^{-0.5})$ [$\sigma_{\rm fit} = (0.47\,N\,(1.1+0.5 I^{-0.9})$].
%
Perhaps the first thing to note in this figure is that, with our adopted methodology, we do not recover the full cross-section of the 81 constituent primary particles but, at most, only about half of that number. Note that the asymptotic limit for the analytical fit to the data (solid orange line) is $\langle \sigma_{\rm eff} \rangle = 1.1(0.52\,N) \ l^2$ or $\approx 57$\% of the cross-section of the infinitely-separated cubic primary particles, i.e., $81l^2$. This arises because we impose the condition that the aggregates must be constructable and contiguous entities, which limits the typical ranges of $D\ (\approx 1.2-2.5)$ and $I\ (\lesssim 5)$ and results in structures with significant shadowing ($\gtrsim 50$\%).  
As shown in the Appendix \S~\ref{sect_averaged} even highly extended aggregates with $D_f = 1$ still shadow some 13\% of the total primary particle cross-section. Note that for such an aggregate our method predicts $\langle \sigma_{\rm eff} \rangle \approx \frac{2}{3}$ and therefore 33\% shadowing, rather than the expected 13\%. In fact only in a system of isolated particles (and therefore by definition not an aggregate) is it possible to recover the cross-section of the infinitely-separated primary particles. This only occurs when there can be no shadowing and the cross-section per primary particle, $\sigma_{pp}$, is equal to that of the total number of particles (i.e., $\sigma_{pp} \approx N\,l^2 = \pi r^2$) and the volume per primary particle is then $\frac{4 \surd \pi}{3}  N^{\frac{3}{2}}\,l^3$, which is equivalent to the value of $I$ where all primary particles are `visible' in the cross-section.  With our methodology we fitted simulations of the cross-section of a `cloud' of 81 randomly-placed and isolated primary particles and find $\sigma = N\,l^2\, (1-1.9 e^{-I^{0.25}})$, which yields $\sigma = N\,l^2$ at an inflation value of $\sim 1700$. 

It is clear from Fig.~\ref{fig_sigma_xyz_vs_I} that $\langle \sigma_{\rm eff} \rangle$ increases with increasing $I$. This is due to what might be considered an `optical depth effect' because the more inflated particles expose more of their constituent primary particles and therefore add to the projected area.  
It also appears that the effects of $I$ on $\langle \sigma_{\rm eff} \rangle$ are stronger than those due to $D$, which perhaps appears at odds with previous studies of soot particles \citep[e.g.,][see Appendix \ref{sect_averaged} for the details]{koylu95}. However, such studies generally only characterise the particles in terms of their fractal dimension $D_f$, whereas here we use a dual characterisation with $D$ and $I$, where $D$ is not equivalent to $D_f$. Hence, a direct comparison of the respective analyses is not really possible. 

The data in Fig.~\ref{fig_sigma_xyz_vs_I} show a `yellow spur' in the upper left region at $\langle \sigma_{\rm eff} \rangle/N = 0.44-0.51$ and $I \lesssim 3$. An analysis of the aggregate shape distribution shows that the yellow spur particles corresponds to `pancake-like' particles with holes in which one of the dimensions $N_x$, $N_y$ or $N_z$ is equal to that of a single primary particle. Such particles are not numerous in this simulation, and are not particularly physical, and we therefore ignore them here.


\subsection{Irregular particle mean surface-to-volume ratios} 

\begin{figure}
\centering 
\includegraphics[width=8.5cm]{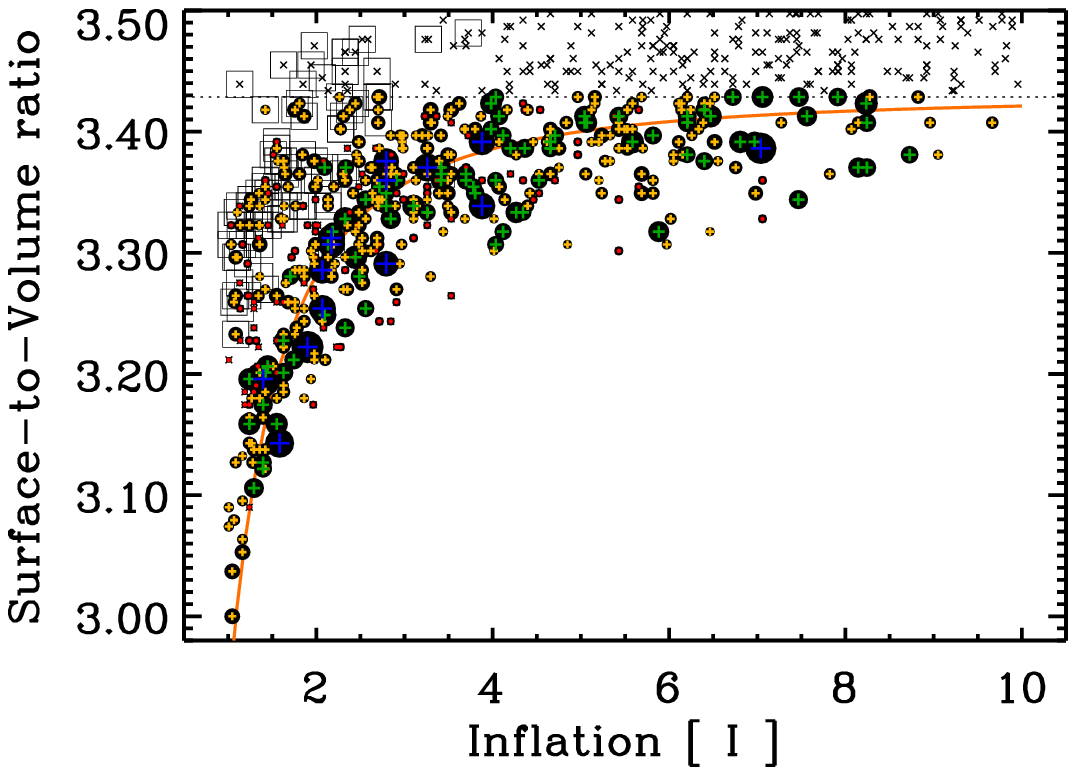}
\caption{The surface-to-volume ratio for $\approx 500$ valid aggregates (filled symbols) with 81 cubic primary particles, on a cubic grid, normalised to that of the most compact `spherical' structure. The filled symbols are size- and colour-coded as per Fig.~\ref{fig_sigma_xyz_vs_I}. The `bare' crosses indicate non-constructable aggregates, as per Fig.~\ref{fig_D_vs_I_limited}.}
\label{fig_surfvol_vs_I}
\end{figure}

The formation of H$_2$ is a critical step, and a key starting point for, interstellar chemistry. Given that its formation occurs principally by H atom (re)combination on grain surfaces an aggregate surface-to-volume ratio, $S_V$, is of importance to interstellar chemistry and we have therefore calculated this ratio for our cubic aggregates.  $S_V$ depends upon the mean {\em facet} coordination number, ${\bar m_{\rm F}}$ ($0 \leqslant {\bar m_{\rm F}} \leqslant 6$), which is the mean number of cube faces that coordinate to other primary particle faces and reduce the available surface. For the most compact `spherical' form with 81 primary particle cubes, and a projected cross-section of $21l^2$ along each axis, we have $S_V = (6 \times 21 l^2/81 l^3) = (126/81l) = 1.56/l$. The maximum possible value, where the cubes are connected only by their edges and all surfaces are therefore exposed, is $S_V = (81 \times 6l^2/81l^3) = 6/l$ (we recall that cube apex-only connections are not allowed here).
For the constructed aggregates we then have
\begin{equation}
S_V \approx \frac{ 81 \times ( 6 - {\bar m_{\rm F}})\, l^2}{81\, l^3} = \frac{ ( 6 - {\bar m_{\rm F}})}{l}, 
\label{ eq_SV0}
\end{equation}
which gives the maximum value when ${\bar m_{\rm F}} = 0$ and the minimum value when ${\bar m_{\rm F}} = 4.44$. 
For each cube the maximum possible coordination number is the sum of the edge (12) and face (6) connections, i.e., 18. Face or {\em facet} connections therefore make up one third of the possible coordination sites and we have ${\bar m_{\rm F}} \approx {\bar m}/3$, which yields   
\begin{equation}
S_V \approx \frac{( 6 - {\bar m}/3)}{l}, 
\label{ eq_SV1}
\end{equation}
and normalising to the minimum surface-to-volume ratio for the most compact structure gives  
\begin{equation}
S_{V, norm} \approx \frac{ 81 \times ( 6 - {\bar m}/3)}{126} = 0.643 \left( 6 - \frac{{\bar m}}{3} \right). 
\label{ eq_SVn}
\end{equation}
In Fig.~\ref{fig_surfvol_vs_I} we show this normalised surface-to-volume ratio for the same aggregates as for Fig~\ref{fig_D_vs_I_limited} but also include the non-constructable aggregates (crosses). 
The horizontal dotted line indicates the clear cut-off between constructable (${\bar m} \geqslant 2$) and non-constructable (${\bar m} < 2$) aggregates, i.e., $S_{V, norm} = [ 81 \times ( 6 - \frac{2}{3})]/126 = 3.43$. 
For the constructable aggregates the mean coordination numbers range between 2 and 4, corresponding to a rather narrow range in $S_V$, i.e., $3.00 - 3.43$. 
Thus, these aggregates do not exhibit large variations in $S_V$, which will also be the case for aggregates of spherical primary particles. 
The points enclosed by squares are the rather unlikely pancake-like, `yellow spur' aggregates in Fig~\ref{fig_D_vs_I_limited} and we again ignore them here.
Also shown is an illustrative empirical `fit' to the data $S_V^\prime  = \{max[S_{V, norm}] - \frac{1}{2}I^{-\frac{7}{4}} \} = (3.43 - 0.5/I^{1.75} )$. 
%
%


\subsection{Applicability to `real' aggregates?}

As interesting as the above results, for the effective cross-sections of aggregates of cubic grains on a cubic lattice, may appear we really need to consider their relevance to `real' aggregates consisting of a size distribution of  `spheroidal' primary particles. 
Firstly, for spheres on the same cubic grid the effective cross-section would be lower by a factor $\pi (l/2)^2 / l^2 = \frac{\pi}{4}$ when we take a mean value based upon views exactly along the cartesian axes of the aggregates, which is a special case that maximises the view of the `pores' between the primary particles.  If we imagine taking the mean values off-set by only a small angle then we quickly `lose sight of' the pores and approach the value of $\langle \sigma_{\rm eff} \rangle$ for the array of cubic primary particles. For spheres on an fcc lattice our derived $\langle \sigma_{\rm eff} \rangle$ is also probably a good approximation because, in the fcc lattice where $\phi = 0.74$ (c.f. $\phi = 1$ for cubic primary particles) and the maximum coordination number is 12, there is intrinsically significant shadowing and in projection the `pore dilution' of $\langle \sigma_{\rm eff} \rangle$ will be negligible. Similarly, if we consider pore filling by a mantling `glue' and/or abundant smaller primary particles, then our `simple cubic cross-section' model probably does a rather good job of estimating the projected cross-section for an aggregate as a function of $D$ and $I$. 

This study is not intended to be exhaustive but is meant to be rather indicative and to give some general guidelines for studies of irregular aggregates. Clearly the exact forms of the particles in a study, and therefore their distributions in the $D$ and $I$ space, will depend on the aggregate construction protocol. Nevertheless, the  `simple cubic cross-section' model is probably a very good approximation and ought to be widely applicable.


\subsection{Implications for gas-grain coupling}

In dynamical studies of porous/irregular grains, where gas-grain coupling plays a key role, it is important to have a good measure of the geometrical, or projected, cross-section of the particles. This is particularly so for porous/aggregate dust-gas coupling leading to dust drag/acceleration in circumstellar and disc winds, and in turbulent or shocked gas. 
Simply `inflating' the grains, using a lowered effective density to model porosity \citep[e.g.,][]{jones94} is probably not sufficient because the inflated cross-section is not filled and the surface-to-mass ratio will lower than that predicted by the simple grain-`inflating' approach. 
In gas-grain coupling studies the projected cross-section for irregular (and `porous') grains, and the appropriate orientation-averaging, therefore needs to be carefully considered as it will depend upon the adopted aggregate construction protocol.


\section{Concluding remarks}
   
We underline the fact that care needs to exercised in the use of the terms porosity and fractal dimension when considering finite-sized interstellar, interplanetary or cometary particles. In detail these grains are not fractal in the true sense of the definition of the term fractal because their structures are not scale-invariant, i.e., a shift in the reference point will not always yield the same fractal dimension. 

Often, in coagulation simulations and experiments, the structures appears  as rather psuedo-centro-symmetric, in the sense that they appear to have a centre from which extended structures radiate. In contrast, IDP grains are rather `blocky' and do not appear to exhibit `radial'-type structures extending away from a centre.

A new means of characterising the averaged `porosity' and `fractality' of irregular particles is proposed.  Given that these two properties of an aggregate are not independent of one another, it seems that may be we should rather think in terms of defining our aggregate, `porous' or `fractal' particles in terms of dimensionality and inflation.
A particle's inflation, $I$, is a measure of its size compared to the most compact form possible and its dimensionality, $D$, is a measure of the 3D disposition of the constituent matter in the  particle. The parameters $I$ and $D$ can be derived from direct measurements of the particle dimensions (normalised to that of the constituent sub-grains in the case of mono-disperse aggregates). 
Their usefulness is, in particular, underlined by the fact that $I$ and $D$  can be applied to particles of any arbitrary shape or structure be they solid and homogeneous, have significant concavities and protrusions or consist of porous aggregates of sub-grains or primary particles.

We find that the cross-sections of irregular aggregates can be rather well approximated by simply modelling them as sparse lattices of cubic primary particles on a simple cubic grid and then averaging along the orthogonal $x$, $y$ and $z$ axes to obtain the mean cross-section. 

This study has focussed on aggregates composed of single-sized sub-grains in an attempt to explore their nature and to propose new measures to allow their characterisation. 
Interstellar grains will in any case be less porous than the rather idealised aggregates considered here because of the effects of ice mantle accretion and the fact that the coagulated grains are formed from a size distribution of primary particles with a clear excess of small over big grains. These factors will tend to lead to pore filling and to aggregates that are more compact and much more strongly bound together due to the extensive surface-to-surface contacts.

A definition of $I$ and $D$ for aggregate particles derived from a size distribution of primary particles will be rather more complex than the simple derivations presented but could, in principle, be derived in an analogous way. 
However, this may not be necessary because the quantities dimensionality, $D$, and inflation, $I$, can be derived from direct measurements of the particle dimensions ($L_i$, where $i = x, y\ {\rm or}\ z$) and the volume (or mass) of the constituent solid matter, $V_s$, i.e.,
\begin{equation}
D = \frac{L_x + L_y + L_z}{{\rm max}\{ L_x, L_y, L_z \}}, 
\end{equation}
and
\begin{equation}
I =  \frac{\frac{1}{6} \pi (L_x \times L_y \times L_z)}{V_s}.
\end{equation}

\begin{acknowledgements}
I would like to thank Melanie K\"ohler and Vincent Guillet for a careful reading of the manuscript and for stimulating discussions on this subject.  
I would also like to thank the referee for an encouraging review and for useful suggestions. 
This research was made possible through the key financial support of the Agence National de la Recherche (ANR) through the program {\it Cold Dust} (ANR-07-BLAN-0364-01).
\end{acknowledgements}

\newpage

\appendix

\section{Extended particle properties}
\label{appendix}

With the interpretation of the interstellar dust emission observations from the Herschel and Planck missions very much in mind, we are obviously interested in the long wavelength emission from dust and dust aggregates, and therefore in the absorption cross-section, which is equal to the emission cross-section, $C_{\rm abs} = C_{\rm em}$. In the following we explore the cross-sectional properties of aggregates of sub-grains.


\subsection{Orientation-averaged cross-sections}
\label{sect_averaged}

The orientation-averaged cross-section factor, $\langle S \rangle$, (the effective cross-section normalised to the maximum cross-section) for a particle is obviously $\langle S \rangle = 1$ for a sphere (all particle axes equal). For a thin disc  \cite{micelotta09} find $\langle S \rangle = \frac{1}{2}$. For a randomly-tumbling cylinder \cite{brown05} show that the effective cross-section is $\frac{1}{2} \pi r ( r + l )$, where $r$ is the cylinder radius and $l$ is its length, which corresponds to:
\begin{equation}
\langle S \rangle = \frac{\frac{1}{2} \pi r ( r + l )}{2\, r\, l} = \frac{\pi}{4}\left( \frac{r}{l} + 1 \right)
\label{Seqn}
\end{equation}
If we take the particle enclosing volume, $V_{\rm e}$, normalised to the sub-particle size $d$,  to be $V_{\rm e} / d^3 = \pi \,( N_x \, N_y \, N_z ) / 6$ (see Eq.\ref{Venc}), then the sub-particle size-normalised, orientation-averaged, `enclosing' particle cross-section is given by:
\begin{equation}
\frac{\sigma_{\rm e}}{d^2} = \langle S \rangle \frac{\pi}{4} ( N_x \, N_y \, N_z )^\frac{2}{3} = \langle S \rangle \frac{\pi}{4} (\Pi^i)^\frac{2}{3}, 
\end{equation}
which is not filled, except for compact particles. However, we need to determine the actual geometrical cross-section of the extended, porous particle (the fraction of the cross-section $\sigma_{\rm e}$ that is filled), i.e., the total projected cross-section, or the geometrical filling factor of \cite{ormel09}. Unfortunately, \cite{ormel09} and \cite{ormel07} give no indication as to how the geometrical filling factors and enlargement parameters were actually calculated for their particles. 

\begin{figure}
\centering 
\includegraphics[width=8.5cm]{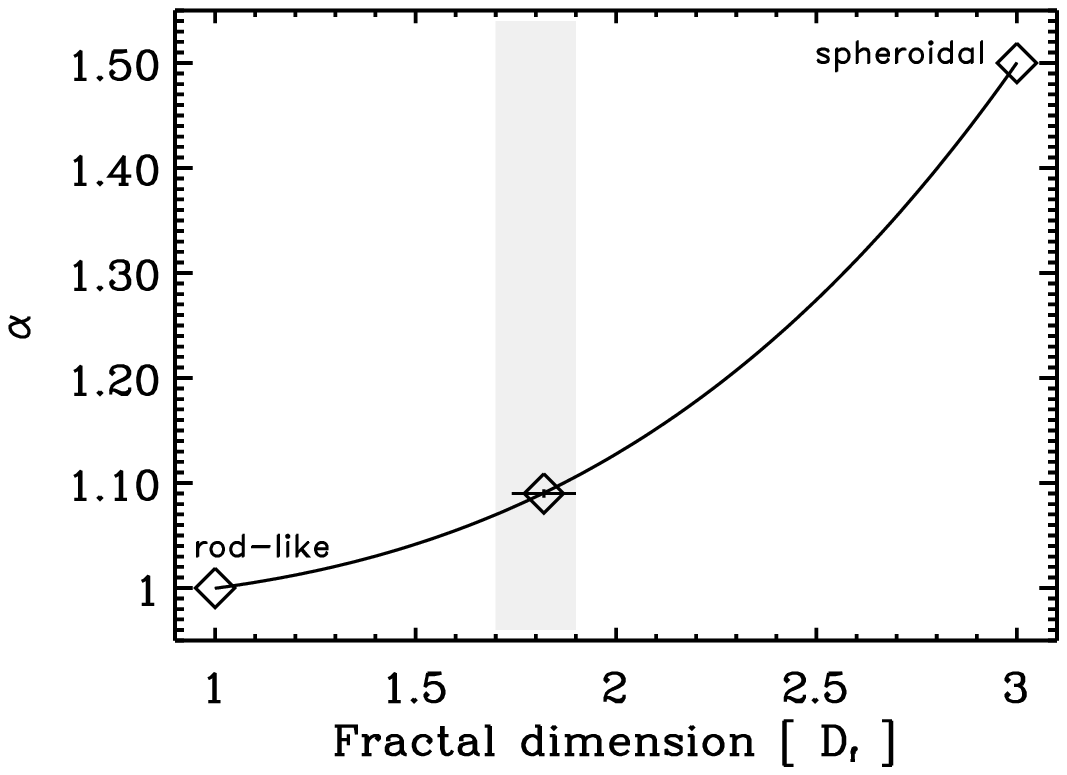}
\caption{Aggregate projected area exponent, $\alpha$, versus fractal dimension, $D_f$, or equivalently the dimensionality, $D$. The data point at $D_f =1.82$ (with the uncertainties in $D_f$ and $\alpha$ indicated by the horizontal and vertical bars) is the best-fit value of $\alpha$ for soot aggregates, which have $1.7 \leq D_f \leq 1.9$ (shaded range), taken from \cite{koylu95}.}
\label{alpha_Df_plot}
\end{figure}

\begin{figure}
\centering 
\includegraphics[width=8.5cm]{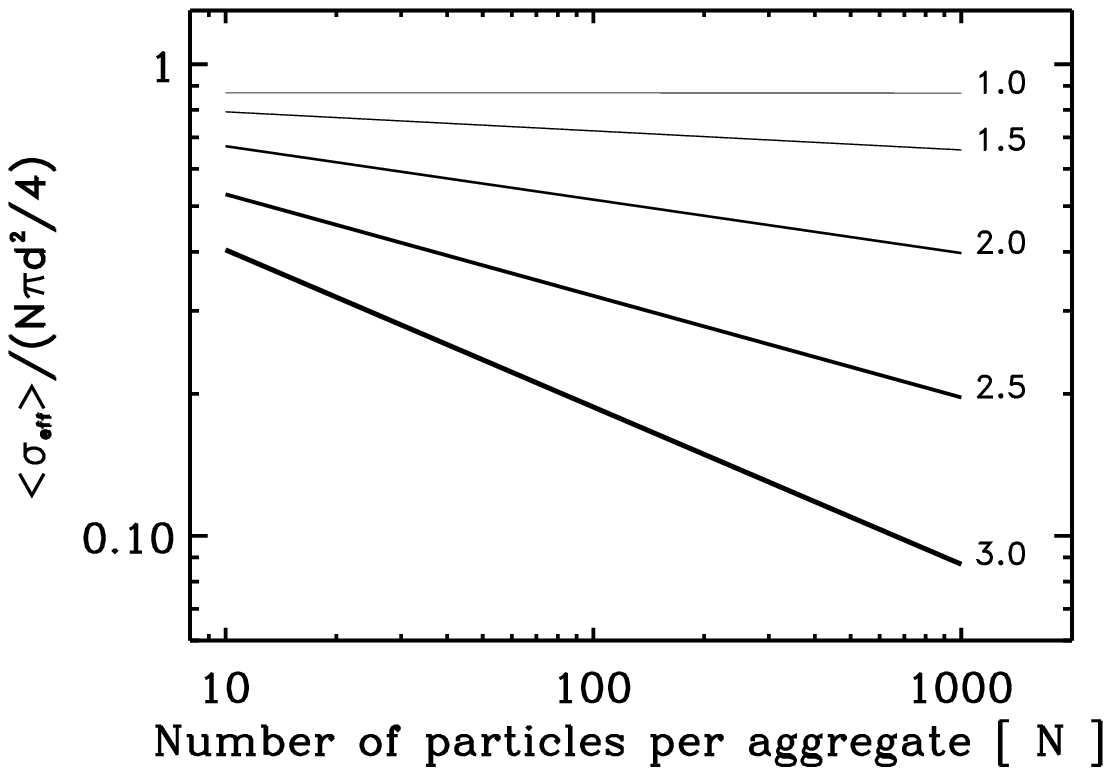}
\caption{The normalised projected cross-section, $\langle\sigma_{\rm eff}\rangle / ( ( N \pi d^2 ) / 4)$, for aggregates consisting of $10-1000$ primary particles of given fractal dimension or fractality (indicated by the number at right). The experimental uncertainties are of about the same order as the thickness of the $D_f = 3$ line.}
\label{sigma_Nnorm_plot}
\end{figure}

This problem was studied by \cite{koylu95}, based on earlier work cited therein,  within the context of soot particle aggregate structures. They found that, for agglomerated soot particles with a fractal dimension $D_f \sim 1.8$, the number of primary particles, $N$, within the aggregate can be expressed as:
\begin{equation}
N = k_a \,( A_a / A_p )^\alpha,
\label{NkAA}
\end{equation}
where $k_a$ is constant close to unity, $A_a$ is projected area of the aggregate and $A_p$ is the cross-section of a primary particle, $\pi (d/2)^2$.  \cite{koylu95} considered aggregates of primary particles or sub-grains. 
A power-law correlation is found to give an excellent fit to their data with $k_a = 1.15\pm0.02$ and an index $\alpha = 1.10\pm0.004$. The experimental uncertainty in $\alpha$ is apparently very small (see Fig~\ref{alpha_Df_plot}) and we will therefore subsequently ignore it. Thus, the projected area, or effective cross-section, $\langle \sigma_{\rm eff} \rangle$, of their fractal particles is given by
\begin{equation}
A_p = A_a \left( \frac{N}{k_a} \right)^{\frac{1}{\alpha}} = \frac{\pi d^2}{4} \left( \frac{N}{k_a}\right)^{\frac{1}{\alpha}} = \langle\sigma_{\rm eff}\rangle.
\label{eq_sigma_proj}
\end{equation}
As noted above this appears to be valid for a limited range of fractal dimension applicable to soot aggregate particles ($D_f \approx 1.7-1.9$). The above expression can be more generally expressed as:
\begin{equation}
\langle \sigma_{\rm eff} \rangle = c N^\beta \times \frac{\pi d^2}{4}.
\end{equation}
For an extended, rod-like aggregate ($D_f =1$), where most, if not all, of the primary particles are `visible', we have $\langle \sigma_{\rm eff} \rangle = ( N \, \pi d^2 ) / 4 $ (i.e., $c=1$ and $\beta =1$), and for the case of the most compact spherical aggregate ($D_f =3$) of primary particles it can be shown that $\langle \sigma_{\rm eff} \rangle = ( N^{\frac{2}{3}} \, \pi d^2 ) / 4 $ (i.e., $c=1$ and $\beta = \frac{2}{3}$). 
Fig.~\ref{alpha_Df_plot} shows these three determined values for $\alpha$ as a function of $D_f$ along with an analytical fit to these data ($\alpha  = 0.984 + 0.0162 D_f^{3.15}$). 
Here, and in the absence of evidence to the contrary, we make the (probably incorrect) assumption that $k_a$ is indepdent of $D_f$ over the entire range of $D_f$.
The use of Eq.~\ref{eq_sigma_proj}, and the dependence of $\alpha$ on $D_f$, now enables us to calculate the projected cross-section for any collection of primary particles of a given fractal dimension or fractality, i.e.,
\begin{equation}
\langle\sigma_{\rm eff}\rangle = 0.870\pm0.015  \ N^{(1 / \alpha)} \ \left( \frac{\pi d^2}{4} \right).
\end{equation}
This expression is plotted in Fig.~\ref{sigma_Nnorm_plot}, as the normalised function $\langle\sigma_{\rm eff}\rangle / ( ( N \pi d^2 ) / 4)$, for aggregates with $10-1000$ primary particles. 
%
%
What this figure shows is that for highly fractal particles ($D_f =1$) the effective cross-section, $\langle\sigma_{\rm eff}\rangle$, scales directly with the number of the primary particles, $N$, and their cross-section, $\pi (d/2)^2$. 
This shows that for $D_f =1$ the angle-averaged cross-section is only 87\% of the total cross-section of all the constituent sub-grains, i.e., $N \pi (d/2)^2$,  which implies that some of the primary particles are `shadowed' in the orientation-averaged projection as clearly must be the case. For a randomly tumbling cylinder ($D_f \sim 1$) setting $\langle S \rangle = 0.87$ in Eq.~\ref{Seqn} corresponds to a cylinder with $l \approx 10 r$. 

As the particle fractality (and hence the `porosity') increase Fig.~\ref{sigma_Nnorm_plot} shows that the normalised effective cross-section decreases rapidly with $N$ as the number of primary particles that are `hidden' within the structure increases. This is simply the effect of decreasing surface-to-volume ratio as the size, or number of primary particles, increases. For typical soot, cluster-cluster aggregates with $D_f \sim 1.7 - 1.9$ it can be seen that the normalised cross-section is $\approx 40-80$\% of the total primary particle cross-section ($N \pi (d/2)^2$).


\subsection{Extended particle radii of gyration}

As pointed out by \cite{mandelbrot82,dobbins91,ossenkopf93}, \cite{koylu95} and \cite{mackowski06}, in aggregated particle studies a quantity of prime importance for the optical and collisional properties of the particles is the radius of gyration, $R_g$, which in these works is defined as:
\begin{equation}
R_g^2 = \frac{1}{N} \sum_i r_i^2, 
\end{equation}
where $r_i$ is the distance of the individual primary particle from the centre of mass of the aggregate. The soot particle and simulation studies by \cite{koylu95} further indicate that
\begin{equation}
N = k_f  \left( \frac{R_g}{d} \right)^{D_f} \ {\rm and \ therefore} \ \ \ \ R_g = \left( \frac{N}{k_f} \right)^{\frac{1}{D_f}} d, 
\label{NkRgd}
\end{equation}
where, from experiment, $D_f = 1.82\pm0.08$ and $k_f = 8.5\pm0.5$. 

\cite{mackowski06} in a similar study, but based only on simulations (references therein), find that 
\begin{equation}
N = k_0  \left( \frac{R_g}{(d/2)} \right)^{D_f} \ {\rm and \ therefore} \ \ \ \ R_g = \left( \frac{N}{k_0} \right)^{\frac{1}{D_f}} d, 
\end{equation}
and that the variation of the factor $k_0$ with $D_f$ can be expressed as $k_0 = (1.27 - 2.2 [D_f -1.82])$.  We can see that the two $k$ factors above are simply related by $k_0 = (1/2)^{D_f} k_f$. Upon substitution it can be shown that there is an apparent discrepancy between the experimental and simulation fits, in that $k_f / k_0 = 1.9$ at $D_f = 1.82$. Nevertheless, we can use the above expressions for $k_0(D_f)$ to derive an equivalent fractal dependence for $k_f(D_f)$, i.e., $k_f = 8.5 (2.82 - D_f)$, which obviously reduces to the above value of $k_f = 8.5$ when $D_f = 1.82$.

In these studies another characteristic, or rather characterising, parameter is the longest particle dimension, the maximum projected length $L$, which is simply related to the above-used quantity, $N_{L+}$ (see \S~5.2), by $L = N_{L+} \times d$. \cite{koylu95} give the following useful relationship:
\begin{equation}
N = k_{fL} \left( \frac{L}{d} \right)^{D_f} = k_{fL} \ N_{L+}^{D_f}, 
\end{equation}
which, when the constant $k_{fL}$ is known, can be used to derive the fractal dimension of a particle from measurements of its number of constituent primary particles, its maximum projected length and the size of the primary particles, i.e.,
$D_f = ( {\rm log}\ N - {\rm log}\ k_{fL} ) / {\rm log}\ N_{L+}  $.

Using the above equations that express $N$ as a function of $A_a$ (Eq.~\ref{NkAA}) and of $R_g$ (Eq.~\ref{NkRgd}), and substituting, we obtain the following relationshipfor $R_g$ as a function $A_a$,
\begin{equation}
R_g = d \left( \frac{k_a}{k_f} \right)^{\frac{1}{D_f}} \left( \frac{A_a}{\pi (d/2)^2} \right)^{\frac{\alpha}{D_f}} .
\end{equation}
This is useful because it allows us to calculate the radius of gyration, $R_g$, of an aggregate from known quantities and the calculated projected cross-section of the particle (Eq.~\ref{NkAA}). However, its utility does depend on knowing, or assuming, values for the parameters $k_a$ and $k_f$.


\subsection{Extended particle optical properties}

The optical properties of soot aggregates (with $D_f \sim 1.8$) have been studied by \cite{dobbins91} who derive the angle- and distribution-averaged absorption and scattering properties. \cite{dobbins91} find that the absorption cross-section for polydisperse, fractal-like aggregates with $D_f = 1.7 - 1.9$ is:
\begin{equation}
{\bar C_{\rm abs}^A} = \frac{4 \pi \ {\bar n^1} x_p^3 \ E(m)}{k^2}
\end{equation}
where, according to their definitions, ${\bar n^1}$ is the average number of primary particles per aggregate, $N$ (the first moment of their aggregate size probability function, $p(N)$), $x_p = (\pi d)/\lambda \ll 1$, $E(m) = -{\rm Im} [(m^2-1)/(m^2+2)]$ and $k = (2 \pi)/ \lambda$. With the, partial, substitution for the primary particle radius $d = 2 a$ and some re-arrangement it can be shown that, with$x = ( 2 \pi a) / \lambda$, the above absorption cross-section is simply, in the more usual nomenclature, 
\[
{\bar C_{\rm abs}^A} = N \pi d^2 \times \frac{2 \pi a}{\lambda} \bigg\{ - {\rm Im} \left[ \frac{m^2-1}{m^2+2} \right] \bigg\}
\]
\begin{equation}
\ \ \ \ \ \ \ \ = \frac{N \pi d^2}{4} \times 4 x \bigg\{ - {\rm Im} \left[ \frac{m^2-1}{m^2+2} \right] \bigg\}
\end{equation}
which is just the total geometrical cross-section of the primary particles, $N \pi (d/2)^2$, multiplied by the absorption cross-section for a single primary particle in the Rayleigh limit. The angle-averaged scattering cross-section derived by \cite{dobbins91} is more complex, involving the second moment of their aggregate size probability function, $p(N)$, but for the purposes of the long-wavelength emission from aggregates does not concern us here.
This approximation, via the Rayleigh-Gans (RG) theory, is often used in the context of soot particle studies. 

As pointed out by \cite{mackowski06}, the RG approximation, in the Rayleigh-limit, generally under predicts the absorption cross-sections of aggregates of spheres that have refractive indices typical of carbonaceous soots. 
\cite{mackowski06} undertook a detailed theoretical/simulation study of the effects of aggregation on the optical properties of soot particles and arrived an elegant and simple model for predicting their absorption in relation to the properties predicted by the RG model. In essence,  Mackowski finds that, for large aggregates ($N \geq 100$), the RG model results underpredict the relevant cross-sections by as much as a factor of two but that this discrepancy strongly depends on the value of the complex refractive index at the wavelength of interest. For soot aggregates the RG model underestimate is found to be of the order of 10\% at visible wavelengths and a factor of two in the mid-infrared. The reader is recommended to take a close look at the work by \cite{mackowski06} for further details of the proposed simple model methodology.

The optical properties of irregular particles have also been dealt with using a gaussian random particle approach 
\citep{muinonen96a, muinonen96b} and a distribution of form factors (DFF) based on the gaussian random particle approach 
\citep{min06a,min06b,mutschke09}. With these models it is possible to calculate the absorption and scattering of highly irregular particles. 

\cite{koehler11} have studied the effects of refractive index, particle-particle contact and grain mantling on the emissivity of coagulated particles within the astrophysical context. They find that the grain-grain contact surfaces introduced in the discrete dipole approximation (DDA) model for coagulated grains can, depending on the material properties, explain the observed enhancement in the particle emissivity at far-infrared and sub-mm wavelengths.


\section{Multi-component effective media}
\label{sect_EMT_averaged}

\begin{figure}
\centering 
\includegraphics[width=8.0cm]{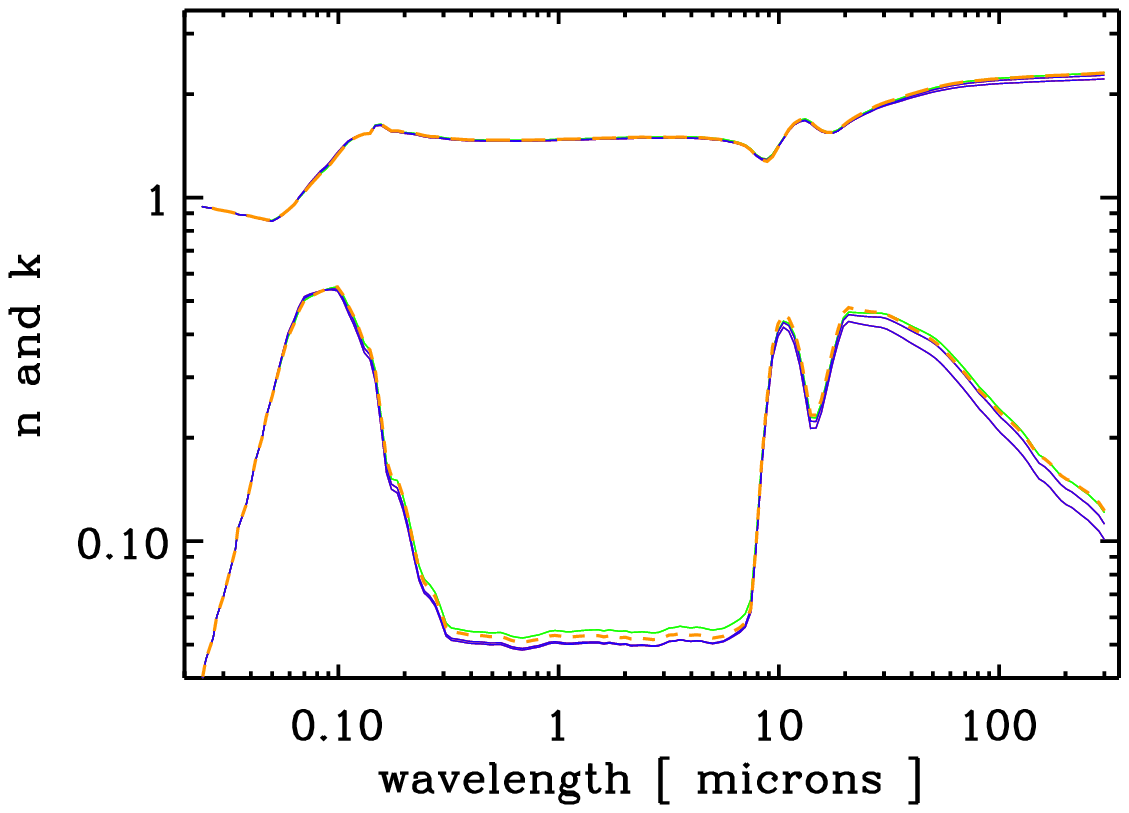}
\vspace*{0.3cm}
\caption{A comparison of the three-component Bruggemann EMT-generated $n$ (upper curves) and $k$ (lower curves) data for a mix of a-Sil (0.4), a-C:H (0.2) and vacuum (0.4) by volume (orange dashed line), with pair-wise mixing using the Bruggemann EMT. 
The pair-wise mixing order [(1+2)$_i$+3$_m$] is: 
[(a-Sil+a-C:H)+vac] and [(a-C:H+a-Sil)+vac]  --  green lines, 
[(a-Sil+vac)+a-C:H] and [(vac+a-Sil)+a-C:H]  --  upper blue  lines, 
[(a-C:H+vac)+a-Sil] and [(vac+a-C:H)+a-Sil]  --  lower blue  lines.}
\label{fig_BEMT3_plot}
\end{figure}

In the Bruggemann effective medium theory (EMT) the effective dielectric function, $\epsilon_{av}$, of a fractional volume of inclusions, $f_i$, of dielectric function $\epsilon_i$ in a matrix of of dielectric function $\epsilon_m$ is found by from the equation \citep{bruggemann35}  
\begin{equation}
f_i \left[ \frac{\epsilon_i - \epsilon_{av}}{\epsilon_i + 2 \epsilon_{av}} \right] + (1-f_i) \left[ \frac{\epsilon_m - \epsilon_{av}}{\epsilon_m + 2 \epsilon_{av}} \right] = 0.
\label{eq_BEMT}
\end{equation}
For a binary mix this involves solving a quadratic equation, for three components a cubic equation etc. In Fig.~\ref{fig_BEMT3_plot} we compare the mixing of the \cite{draine84} astronomical silicate (a-Sil), \cite{rouleau91} BE amorphous carbon (a-C:H) and vacuum components, obtained using the Bruggemann rule for three components, with a pair-wise mixing of the same three components in the same volume fractions. In this pair-wise mixing scheme components $1_i$ (inclusion) and $2_m$ (matrix) are first mixed and then the `effective' result of this pair mixed with a third component, i.e., $[(1+2)_i+3_m]$. We note that in the pair-wise mixing case the results are not commutative, except for the case of switching inclusions and matrix in the first pair-wise effective medium calculation. We find that where the mixing is performed in the order of most refractive to least refractive, i.e., silicate $\rightarrow$ carbon $\rightarrow$ ice $\rightarrow$ vacuum, the pair-wise method gives an excellent match to $n$ and $k$ in comparison with the exact three-component Bruggemann calculation, except for the lowest values of $k$ at mid-infrared wavelengths (see Fig.~\ref{fig_BEMT3_plot}). 
%

In the absence of having solved the quartic, and higher equations, for four and more component mixes we postulate that this pairwise calculation of an effective medium using the Bruggemann mixing rule can be used to calculate the effective medium of any number of mixed media, provided that the materials are mixed in the order of most to least refractory. Thus, this method (the pair-wise Bruggemann EMT or PWB-EMT) can be used to limit the complexity of the calculation of multi-material effective media to the solution of quadratic equations. For example, this method could facilitate studies of inhomogeneous, irregular and `porous' aggregates consisting of amorphous silicates and amorphous carbons with ice mantles, i.e., a four-component mix, without having to solve the quartic version of Eq.~\ref{eq_BEMT}.


\section{$C_{\rm abs}$ at long wavelengths for $n$ constant}
\label{sect_EMT_averaged_longwav}

\begin{figure}
\centering 
\includegraphics[width=8.5cm]{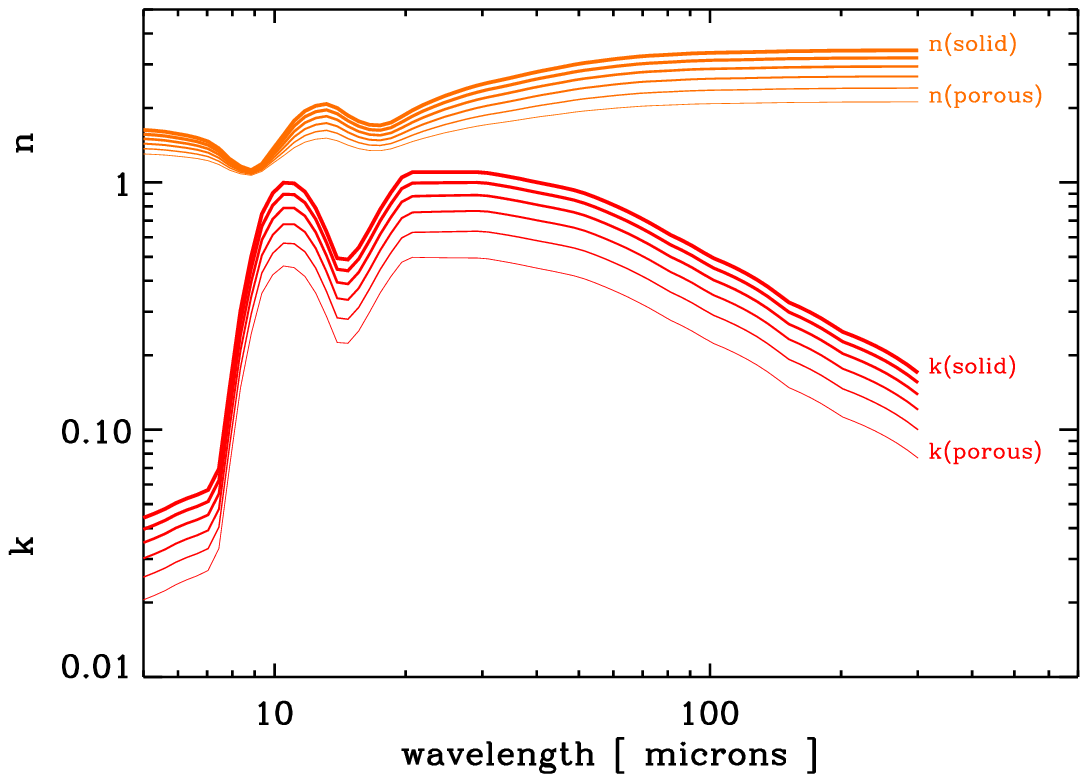}
\vspace*{0.3cm}
\caption{$n$ (upper) and $k$ (lower) for astronomical silicate particles with 0, 10, 20, 30, 40 an 50\% vaccuum (line thickness decreases with increasing vacuum fraction) calculated using the Bruggemann EMT.}
\label{fig_nk_porous_plot}
\end{figure}

At long wavelengths the absorption (and emission) cross-section of a particle of radius $a$ is given by
\begin{equation}
C_{\rm abs} = 4 x \ \bigg\{ - {\rm Im} \left[ \frac{m^2-1}{m^2+2} \right] \bigg\} \ \pi a^2. 
\label{eq_Cabs_1}
\end{equation}
Substituting $x = 2 \pi a_{\rm eff} / \lambda$, where $a_{\rm eff}$ is the effective radius of the solid matter in the aggregate/particle, and $m = n - ik$ for the complex index of refraction, and solving for the imaginary part of the refractive index term we have
\begin{equation}
C_{\rm abs} = 8 \pi \left( \frac{a_{\rm eff}}{\lambda} \right) \ \bigg\{ \frac{6nk}{(n^2-k^2+2)^2 + 4 n^2 k^2} \bigg\} \ \pi a^2. 
\label{eq_Cabs_2}
\end{equation}
For astronomical silicate grains \citep{draine84} at wavelengths longer than 100\,$\mu$m $n$ is constant and $n > k$ ($n \simeq 3.4$, $k \lesssim 0.5$, see Fig.~\ref{fig_nk_porous_plot}, i.e., $n^2 \simeq 11.6$, $k^2 \lesssim 0.25$ and $n^2 k^2 \lesssim 2.9$). 
Neglecting the $k^2$ terms the above yields a simple expression for the absorption cross-section for solid and `porous' \cite{draine84} astronomical silicate grains, i.e., 
\[
C_{\rm abs} =  8 \pi \left( \frac{a_{\rm eff}} {\lambda\ (1-f_v)^{\frac{1}{3}}} \right) \left[ \frac{6n}{(n^2+2)^2} \right] \bigg\{ k_{\rm solid} \bigg\} \  \pi a^2
\]
\begin{equation}
\ \ \ \ \ \ \  \ =   2.79 \left( \frac{a_{\rm eff}} {\lambda\ (1-f_v)^{\frac{1}{3}}} \right) \bigg\{ k_{\rm solid} \bigg\} \ \pi a^2
\label{eq_Qabs_approx}
\end{equation}
where $k_{\rm solid}$ is the imaginary part of the complex index of refraction of the solid material at wavelength $\lambda$ and $f_v$ is the volume fraction of vacuum in the particle. Note that $a$ is the outer radius of the `inflated' porous particle and $a_{\rm eff}$ is the effective radius for the solid matter within the particle. For non-porous particles $ a = a_{\rm eff}$. The above approximation to $C_{\rm abs}$ is equivalent to a simple dilution of the absorptive index, $k_{\rm solid}$, of the solid material by a factor $(a/a_{\rm eff})^3$.  

\begin{figure}
\centering 
\includegraphics[width=9.0cm]{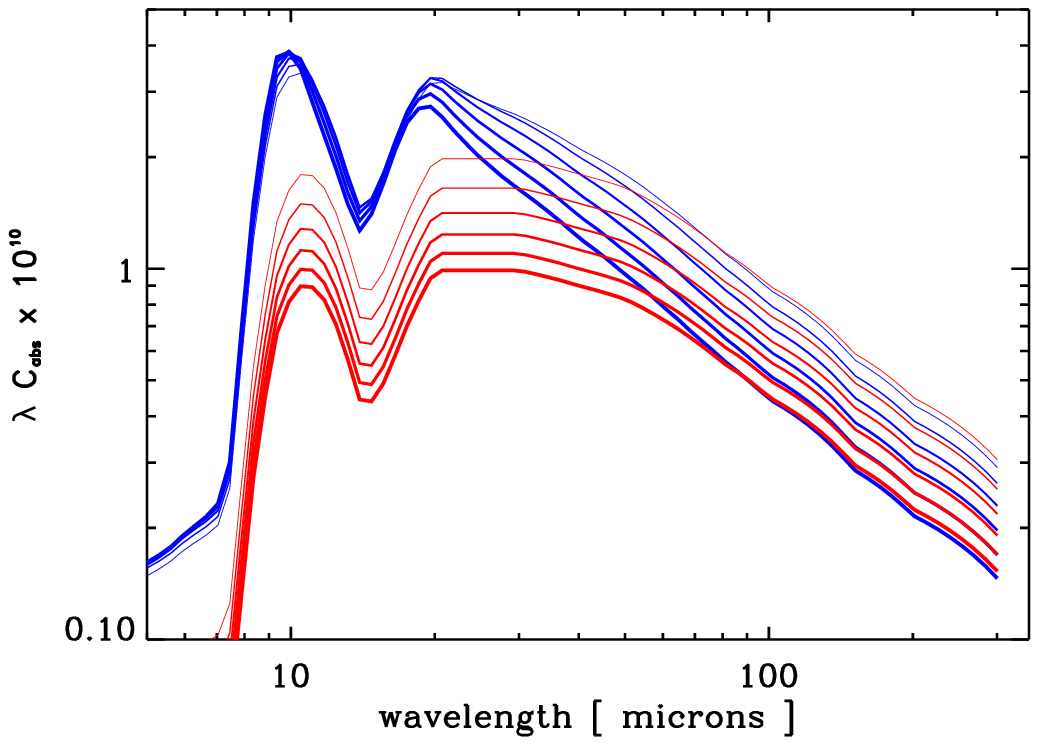}
\vspace*{0.3cm}
\caption{The absorption cross-section, multiplied by the wavelength for clarity, for astronomical silicate particles with 0, 10, 20, 30, 40 an 50\% vaccuum (line thickness decreases with increasing vacuum fraction), and effective radii 100, 104, 108, 113, 119  and 126\,nm, respectively, calculated using Mie theory and the Bruggemann EMT (blue lines). The red lines show the simple analytical fits (see text).}
\label{fig_Cabs_porous_plot}
\end{figure}

In Fig.~\ref{fig_Cabs_porous_plot} we show this fit, together with that calculated using the Mie theory (and the Bruggemann EMT where there is a vacuum component). We find that this approximation can reproduce $C_{\rm abs}$ to within $\pm 6$\% for porosities $\lesssim 50$\%. 
This analytical approximation is only valid at long wavelengths (e.g., the far-infrared and longward) and for materials where $n$ is constant. 
For amorphous carbons in general $n$ is not constant at long wavelengths and this very simple approximation will not work for these materials. 
%
%

This surprisingly-simple approach shows that we need not resort to the Mie theory nor an EMT-averaging to calculate the the dust cross-sections for solid or porous grains at long wavelengths where $n$ is constant.  
However, its usefulness needs to be tested against laboratory data for real interstellar dust analogue materials at long wavelengths.

\end{document}